\documentclass[11pt]{article}
\usepackage{amsmath,amsfonts, amssymb,graphicx,geometry,authblk,color,soul,comment,bm} 
\usepackage[colorlinks=true,linkcolor=black,anchorcolor=black,citecolor=black,filecolor=black,menucolor=black,runcolor=black,urlcolor=black]{hyperref} 

\graphicspath{{figures/}}

\title{A macroscopic constitutive relation for isotropic-genesis, polydomain liquid crystal elastomers}
\author[1]{Victoria Lee}
\author[2]{Adeline Wihardja}
\author[2]{Kaushik Bhattacharya\footnote{Corresponding Author.  Email: bhatta@caltech.edu}}
\affil[1]{Saint-Gobain Competency Research Laboratory, Northborough MA 01532  \footnote{Victoria Lee conducted this work while affiliated with the California Institute of Technology.}}
\affil[2]{California Institute of Technology, Pasadena CA 91125}

\begin{document}
\maketitle
\begin{abstract}
Liquid crystal elastomers (LCEs) are rubber-like solids that incorporate nematic mesogens (stiff rod-like molecules) as a part of their polymer chains.  In recent years, isotropic-genesis, polydomain liquid crystal elastomers (I-PLCEs) has been a topic of both scientific and technological interest due to their intriguing properties such as soft behavior, ability to dissipate energy and stimuli response, as well as the ease with which they can be synthesized.  We present a macroscopic or engineering scale constitutive model of the behavior of I-PLCEs.  The model implicitly accounts for the complex evolution of the domain patterns and is able to faithfully capture the experimentally observed complex response to multi-axial loading.  We describe a multiscale framework that motivates the model, explore various aspects of the model, validate it against experiments, and finally verify and demonstrate a numerical implementation. 
\end{abstract}

\section{Introduction}

Liquid crystal elastomers (LCEs) are rubber-like solids that incorporate nematic mesogens (stiff rod-like molecules) as a part of their polymer chains \cite{warner_2003}.  The cross-linking in these materials is sparse enough that the nematic mesogens retain their liquid crystalline character and associated order-disorder phase transitions.  Importantly, they affect the orientation of the polymer chains as they undergo these phase transitions, and this results in a coupling between temperature, liquid crystalline order and deformation.  This in turn results in complex thermo-mechanical behavior which has motivated a number of applications including actuation \cite{white_2015} and impact resistance \cite{saed_2021}.  The goal of this work is to present a macroscopic or engineering scale model of the behavior of a particular class of liquid crystal elastomers.

LCEs were  envisioned theoretically by de Gennes in 1975 \cite{degennes_1975}, and were first synthesized reliably by K\"upfer and Finkelmann in 1991 \cite{kupfer_1991} (also see \cite{finkelmann_1981} for previous attempts).  These involved nematic side-chains in polysiloxane polymers and primarily monodomain (uniform nematic arrangement) specimens; these are challenging to synthesize.  So, while they enabled to a number of scientific studies and hinted to a variety of applications, actual applications remained limited.   The observation of soft behavior in isotropic-genesis, polydomain LCEs \cite{clarke_1998,fridrikh_1999,urayama_2009}, the development of new chemistries \cite{white_2015,yakacki_2015,saed_2017} and a variety of directed methods of synthesis \cite{ware_2015,ambulo_2017} have made these materials widely available, and the subject of both fundamental and applied studies.  

Various LCEs with nematic, smectic and cholecteric order have been synthesized and studied; see \cite{warner_2003} for a comprehensive discussion.  In this paper, we focus on the most widely studied nematic elastomers, those LCEs with nematic order.   The nematic mesogens are disordered at high temperatures and have nematic order (fluctuating along a common direction or director) at low temperature.  A specimen with a uniform nematic director is a {\it monodomain} specimen, and one with non-uniform (patterned or complex) nematic director pattern a {\it polydomain} specimen.  These materials are typically synthesized by making the polymer and then cross-linking them.  A specimen cross-linked in the isotropic state is said to have {\it isotropic genesis}, and those cross-linked in the nematic state  {\it nematic genesis}.  Isotropic genesis LCEs are macroscopically isotropic as synthesized and develop complex director patterns due to local symmetry-breaking as they undergo the order-disorder transition when cooled.  Thus, they are often referred to as isotropic-genesis, polydomain LCEs (I-PLCEs).  

An important and intriguing property of LCEs is the so-called {\it soft} or {\it semi-soft} behavior.  It was first observed in uniaxial tension in monodomain specimens \cite{kupfer_1994,kundler_1995}: when a specimen with a uniform director orientation is loaded in a uniaxial stress in a direction perpendicular to the director, it shows a soft plateau where the stress remains largely constant at a small value over a large range of strains.  The mechanism of this soft behavior is director reorientation through the formation of stripe domains.  This soft behavior was also observed in polydomain specimens where the directors are not universally distributed \cite{clarke_1998,fridrikh_1999}, and it was shown that it is related to a polydomain-monodomain transition.  It has since been clarified that the the soft behavior is restricted to isotropic-genesis polydomain specimens where the LCE is cross-linked in the isotropic state, and absent in the nematic-genesis LCEs \cite{urayama_2009}.  

All of these investigations were conducted in uniaxial stress.  Recently, it was shown in biaxial stretch experiments \cite{tokumoto_2021}  that the soft behavior in uniaxial stress is a particular manifestation of a broader soft phenomenon labelled {\it in-plane liquid-like behavior}.  When a I-PLCE is subjected to unequal biaxial stretch, there is a regime when the true stress in the two principal directions remain equal (even when the two stretches are unequal).  Further, the true stress depends only on the areal stretch (the product of the two stretches) independent of their individual value or loading history.  In other words, I-PLCE can accommodate shear with any shear stress (up to a point).  Furthermore, wide angle x-ray scattering (WAXS) was used to observe the evolution of the nematic director during loading, and  provided an insight  into mechanisms of the in-plane liquid-like behavior.  Since a I-PLCE is incompressible and isotropic, two independent stretches are sufficient to characterize the material.  Therefore these biaxial stretch experiments provide a comprehensive view of the soft behavior.

The observed soft behavior is also extremely sensitive to loading rate: the stress-strain response in uniaxial tension is observed to stiffen as the loading rate increases from the soft plateau behavior in quasi-statics towards elastic (rubber-like) behavior with a less pronounced and higher plateau at higher rates  (\cite{clarke_1998a, hotta_2001} in siloxane-based side chain I-PLCE and \cite{azoug_2016,linares_2020} in acrylate-based main chain I-PLCE materials).  The linear viscoelastic behavior has also been characterized \cite{hotta_2003,linares_2020}.  All of this work is in uniaxial tension.  Systematic studies in multi-axial loading remain to be performed, though there are measurements of the impact of a spherical projectile on a flat elastomer pad \cite{saed_2021}.  

In addition to an inherent interest in the phenomenon, the soft behavior is also the basis of a number of important applications including biological applications that exploit the resulting variable stiffness \cite{agrawal_2013}, for the control of wrinkling in stretched membranes \cite{plucinsky_2017} and for energy absorption under impact \cite{saed_2021,jeon_2022}.  Further, the soft behavior also leads to extremely high fracture toughness \cite{fan_2016,annapooranan_2022}, and enhanced adhesion \cite{ohzono_2019}.  The progress of all these applications requires an engineering model that has the fidelity to describe the complex phenomenology.  This is the motivation for this work.

Bladon, Warner and Terentjev \cite{bladon_1993} extended the classical approach to modeling elastomers to LCEs and developed the neo-classical theory -- the polymer chains are still described by Gaussian statistics, but in an anisotropic medium. This theory was able to explain the soft behavior and the formation of the stripe domains \cite{verwey_1996}.  Desimone and Dolzmann \cite{desimone_2002} recognized that the neo-classical theory led to an elastic energy density that is not quasi-convex, and this lack of convexity can manifest itself into a vast range of domain patterns (including but not limited to stripe domains).  They computed the relaxation --  the overall energy after the LCE has formed the best adapted domain pattern, and showed that this has a range of degenerate behavior.  The approaches concerned the ideal behavior, and predict an ideal stress plateau at zero stress.  It was recognized that the cross-link density is not uniform \cite{fridrikh_1997} and this led to an extension of the neo-classical model to the non-ideal setting \cite{biggins_2008}.  It remains an open problem to compute the relaxation for this non-ideal model.  However, it is possible to use bounds to characterize the overall behavior: this provides an understanding of the difference between the isotropic and nematic genesis for example \cite{biggins_2009,biggins_2012}.  Further, detailed numerical simulations provide insights into the evolution of the domain patterns and their macroscopic consquences in an I-PLCE under complex loading conditions \cite{zhou_2021}, and specific phenomena \cite{barnes_2023}.  However, these are much too expensive to use in an engineering context.

The dynamics of domain evolution has also been studied extensively phenomenologically (see discussion in \cite{warner_2003} for the difficulties in developing a first principle model of dynamics).  At the level of individual domain patterns, one can adapt the Leslie-Ericksen theory of nematic viscosity \cite{ericksen_1961,leslie_1968} to LCEs \cite{brand_1994,clarke_2001,warner_2003} in the small distorsion setting.  This was revisited recently by Wang {\it et al.} \cite{wang_2022} in the finite deformation setting for monodomain specimens.  While these theories are formulated in three dimensions, they are only fitted to uniaxial tension.  Further, these are limited to monodomain or simple domain patterns.

In this paper, we develop a macroscopic or engineering scale constitutive model of the behavior of isotropic-genesis polydomain liquid crystal elastomers (I-PLCE).  We start from a multiscale framework described in detail in Section \ref{sec:back} with a separation of scales between the specimen, the domain pattern and domain scales.  We then take insights from multi-axial experiments and WAXS observation of the domain pattern evolution \cite{tokumoto_2021} as well as the numerical simulations \cite{zhou_2021} to develop appropriate order parameters or state variables that describe the overall response and dynamics at the specimen or engineering scale.  We present the detailed constitutive model in Section \ref{sec:model}.  It uses the state variables developed earlier to implicitly account at the specimen scale the evolution of the nematic mesogens at the domain patterns and domain scale.  We demonstrate the model against biaxial experiments and then conduct a parameter study to study various aspects of the constitutive model.  We then implement the model in the commercial finite element platform \texttt{ABAQUS} \cite{abaqus} in Section \ref{sec:comp}.  We verify the implementation using biaxial loading, and demonstrate it using a study of torsion.  We conclude in Section \ref{sec:conc} with a discussion.

\section{Multiscale setting} \label{sec:back}

In this section, we describe the multiscale setting of I-PLCEs and the insights it provides for the overall behavior of such materials.  These provide the background and heuristic considerations for the constitutive relation proposed in the next section.  A reader interested only in the constitutive model can skip this section on first reading. 

\subsection{Multiscale setting and order parameters}

\begin{figure}
\centering
\includegraphics[width=4in]{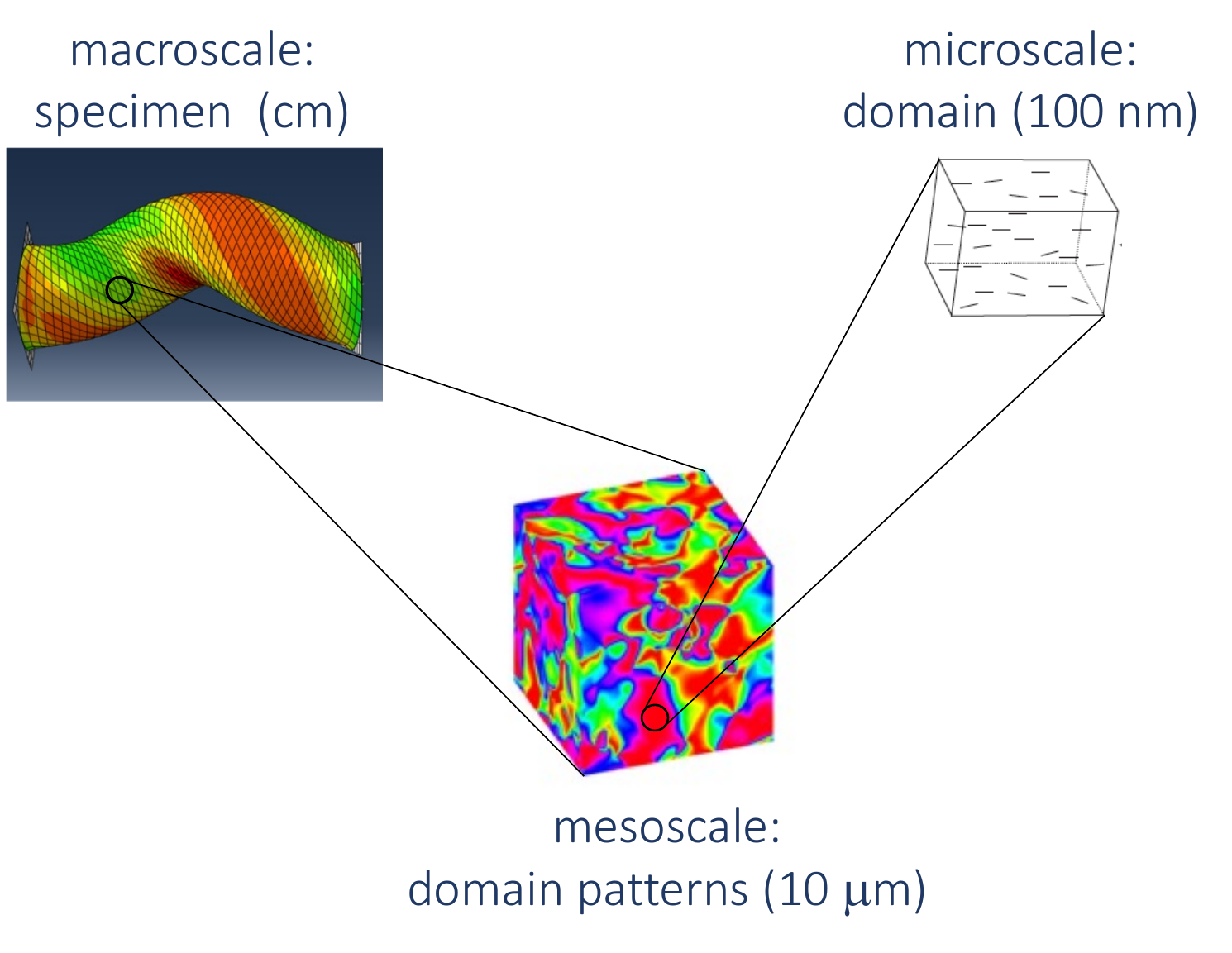}
\caption{A polydomain isotropic-genesis liquid crystal elastomer has a multiscale structure \label{fig:multi}}
\end{figure}

A isotropic-genesis polydomain liquid crystal elastomer (I-PLCE) at a temperature below the isotropic-nematic transition temperature has a multiscale structure shown schematically in Figure \ref{fig:multi} \cite{urayama_2009,zhou_2021}.  We have a specimen at the macroscale or application-scale of millimeters to centimeters.  At the mesoscale of one to ten microns, we have domain patterns.  The domain patterns may be different in different regions of the specimen.  Typically, the domain patterns are not easily characterizable (as for example in stripe domains), but consist of a distinct domains of largely uniform nematic mesogen orientation at a length-scale  the 10-100 nanometers.  

Our goal is to build a model that can be used for engineering calculations of  macroscale or application-scale I-PLCE specimens.   This means that we have to pick a representative volume element (RVE) at the mesoscale and describe the overall or averaged behavior of the domain pattern.  A key descriptor of the domain pattern is the {\it orientation order tensor} that describes the one point statistics of all the mesogens in the representative volume.  We define
\begin{align}
S = \left\langle u \otimes u - {1\over 3} I \right\rangle_\text{mesogens in RVE}
\end{align}
where $u$ is an unit vector describing the orientation of a mesogen and $\langle q \rangle_{\Omega}$ denotes the average of a quantity $q $ over a region $\Omega$.  Since the mesogen is not polarized, the sign of $u$ is not meaningful and therefore it is natural to take the dyadic product.  It is also conventional to subtract a third of the identity to make $S$ trace-free.

It is useful to first average over individual domains, and then over all domains:
\begin{align}
S&= \left\langle \left\langle u \otimes u - {1\over 3} I \right\rangle_\text{mesogens in a domain} \right\rangle_\text{domains in RVE}
\end{align}
Now, focus on the inner average: since the mesogens fluctuate around an average direction within a domain, we may write
\begin{align}
 \left\langle u \otimes u - {1\over 3} I \right\rangle_\text{mesogens in a domain} =  Q \left(n \otimes n - {1\over 3} I \right) 
 \end{align}
where $Q$ is the {\it microscopic degree of order} within a domain, and $n$ is the average orientation or {\it director}.  Note that $Q$ takes values between $0$ and $1$ and depends on temperatures; for a typical LCE at room temperature, $Q \approx 0.6$ \cite{warner_2003}.  Substituting this back, 
\begin{align}
S = Q \left\langle n \otimes n - {1\over 3} I \right\rangle_\text{domains in RVE} 
= Q \hat{S}
\end{align}
where 
\begin{align}
\hat{S} = \left\langle n \otimes n - {1\over 3} I  \right\rangle_\text{domains in RVE}
\end{align}
is the {\it mesoscale orientational order tensor} that describes how the domains are distributed in the RVE.  It differs from the overall 
orientational order tensor by the microscopic degree of order $Q$.

\begin{figure}
\centering
\includegraphics[width=5in]{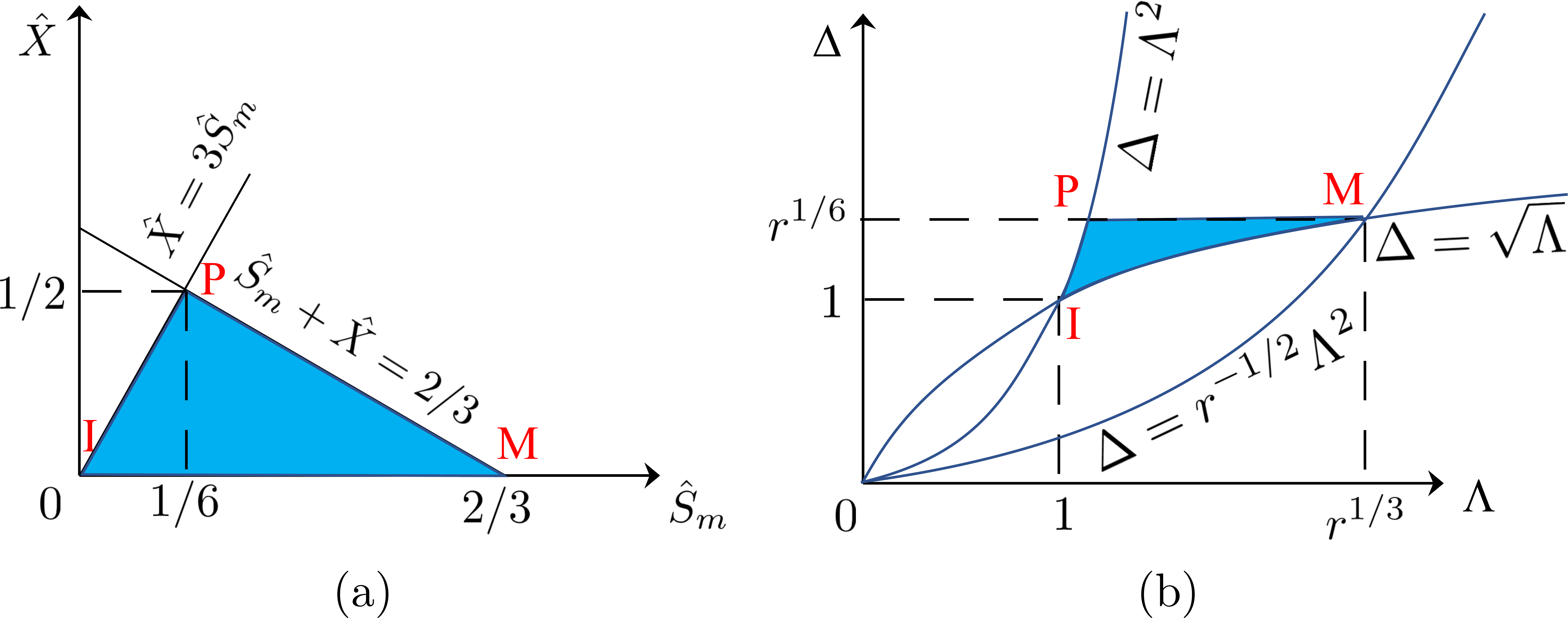}
\caption{The shaded regions indicate the (a) allowable domain order parameters ${\mathcal A}_S$ and (b) allowable spontaneous deformation parameters ${\mathcal A}_G$.    The values of the parameters corresponding to the isotropic $I$, monodomain $M$ and planar equiaxed $P$ are also indicated. \label{fig:order}}
\end{figure}

Now, since $S$ (respectively $\hat{S}$) is symmetric, it has three eigenvalues taking values between $-1/3$ and $2/3$.  Further, since it is trace-free, only two of them are independent.  The largest eigenvalue $S_m$ (respectively $\hat{S}_m$) is the {\it mean orientational order parameter} (respectively mean domain orientational parameter) and describes how the mesogens (respectively directors) align along the principle direction in an RVE.  The difference $X$ (respectively $\hat{X}$) between the two smallest eigenvalues of $S_m$ (respectively $\hat{S}_m$) is the {\it mean biaxial order parameter} (respectively mean domain biaxial order parameter)  It describes the planarity of the mesogen (respectively director) orientation.   Thus, in the principal basis,
\begin{align} \label{eq:s}
S = \begin{pmatrix} S_m & 0 & 0 \\ 0 & {1 \over 2} (-S_m + X) & 0 \\ 0 & 0 &  {1 \over 2} (-S_m - X) \end{pmatrix}
= Q \begin{pmatrix} \hat{S}_m & 0 & 0 \\ 0 & {1 \over 2} (-\hat{S}_m + \hat{X}) & 0 \\ 0 & 0 & {1 \over 2} (-\hat{S}_m - \hat{X}) \end{pmatrix}
= Q \hat{S}.
\end{align}
Table \ref{tab:orient} shows the orientational order parameters for some  domain patterns.  Further the order parameters are limited to the allowable set ${\mathcal A}_S$:
\begin{equation}
(\hat {S_m}, \hat{X}) \in {\mathcal A}_S = \{(\hat {S_m}, \hat{X}): 
\hat{X} \ge 0, \quad \hat{X} \le 3 \hat{S}_m, \quad \hat{X} + \hat{S_m} \le 2/3\}
\end{equation}
to satisfy the constraints on the eigenvalues of $\hat{S}$.  This is shown in Figure \ref{fig:order}(a).

\begin{table}
\centering
\caption{Orientational order parameters for some domain patterns \label{tab:orient}}
\begin{tabular}{|c|c|c|c|}
\hline
Domain pattern &Isotropic (I)&Monodomain (M)&Planar (P)\\
\hline
$S_m$ & 0& $2Q/3$ & $Q/6$\\
\hline
$X$ & 0 & 0 & $Q/2$\\
\hline
$S$ (principal axes) & 
$  \begin{pmatrix} 0 & 0 & 0 \\ 0 & 0 & 0 \\ 0 & 0 & 0 \end{pmatrix}$&
$ Q \begin{pmatrix} \frac{2}{3} & 0 & 0 \\ 0 & -\frac{1}{3} & 0 \\ 0 & 0 &  -\frac{1}{3} \end{pmatrix}$ &
$ Q \begin{pmatrix} \frac{1}{6} & 0 & 0 \\ 0 & \frac{1}{6} & 0 \\ 0 & 0 &  -\frac{1}{3} \end{pmatrix}$\\
\hline
\end{tabular}
\end{table}

\subsection{Spontaneous Deformation} \label{sec:sd}

We begin at the microscale.  According to the neo-classical theory of Bladon, Warner and Terentjev \cite{bladon_1993}, a domain has a spontaneous deformation with Cauchy-Green (metric) tensor $\ell$ that is uniaxial with a stretch along the director $n$ and volume preserving:
\begin{equation} \label{eq:sl}
\ell = r^{-1/3} ( I + (r-1) n \otimes n)
\end{equation}
where $r$ depends on the microscopic degree of order $Q$.  In the case of a freely jointed rod model, $Q = (r - 1)/(r + 2)$ \cite{bladon_1993}, and therefore $r \approx 8$ in a typical I-PLCE at room temperature.  Turning now to the mesoscale, we analogously associate a state of spontaneous deformation with right Cauchy-Green (metric) tensor $G$ with each allowable domain pattern.  Since $G$ is positive-definite and symmetric, we may write 
\begin{equation}
G = P G_0 P^T
\end{equation}
where $P$ is a rotation and $G_0$ is diagonal with positive entries in a fixed laboratory frame.  Further, we assume that the LCE is incompressible, and consequently, the spontaneous deformation is isochoric: hence, $G_0$ has two independent eigenvalues.  For future use, we use the principal stretch $\Lambda$ and the principal areal stretch $\Delta$: so
\begin{equation} \label{eq:g0}
G_0 = 
\begin{pmatrix} \Lambda^2 & 0 & 0 \\ 0 & \Delta^2/\Lambda^2 & 0 \\ 0& 0 & 1/\Delta^2 \end{pmatrix}
\end{equation}
with $\Lambda \ge \Delta/\Lambda \ge 1/\Delta$.  We call $\Lambda, \Delta$ the {\it descriptors of spontaneous deformation}.  
The spontaneous deformation is related to the domain pattern.  Comparing equations (\ref{eq:s}) and (\ref{eq:g0}), we can identify $\Delta$ with $S_m+X$, and $\Lambda$ with $S_m$.  

There are limits on the values that the descriptors of spontaneous stretch can attain.  In the monodomain state, the spontaneous deformation is such that  $G= \ell$.  Further, for all domain patterns, it follows from the convexity of the the principal stretch and principal areal stretch, that the values of descriptors of spontaneous stretch are limited by those in the monodomain state.  Therefore, we have
\begin{equation} \label{eq:gell}
\Lambda \le r^{1/3},  \quad \Delta \le r^{1/6}.
\end{equation}
We also have the limitation $\Lambda \ge \Delta/\Lambda \ge 1/\Delta$.  It follows that, the principal stretch and principal areal stretch are limited to the set ${\mathcal A}_G$:
\begin{equation} \label{eq:ag}
(\Lambda, \Delta) \in {\mathcal A}_G = \{(\Lambda, \Delta): \Delta \le r^{1/6}, \Delta \le \Lambda^2, \Delta \ge \sqrt{\Lambda}\}.
\end{equation}
This set is shown in Figure \ref{fig:order}(b).

\subsection{Free energy density}

\paragraph{Domain scale}

The free energy density of a nematic liquid crystal at the domain scale is given by \cite{warner_2003}
\begin{equation}
{\mathcal F}^\mu(F,n,\nabla n,x, Q,T) =  {\mathcal F}^\mu_e(F,n,r(Q),T) + {\mathcal F}^\mu_{ni}(n,x) + {\mathcal F}^\mu_{LC}(Q,T)+{\mathcal F}^\mu_{F}((\nabla n)F^{-T})
\end{equation}
$F$ is the deformation gradient relative to the isotropic reference scale, $n$ the director, $\nabla$ the reference gradient operator, $Q$ the microscopic degree or order and $T$ the temperature.  The terms in order are the entropic elasticity of the rubber chain, the ``non-ideality'' introduced by disorder in the cross-links, the energy due to steric and enteric energy of the nematic mesogens and finally the Frank elasticity that promotes spatial homogeneity in the nematic mesogen.

It is natural to minimize $Q$ out of the problem holding the other variables fixed.  The liquid crystal energy ${\mathcal F}^\mu_{LC}$ (which may be modeled phenomenologically using a Landau polynomial or more fundamentally using Maier-Saup\'e molecular theory) is the dominant term, and minimizing this gives us the order parameter as a function of temperature, $Q = Q(T)$.  Thus, the free energy is,
\begin{equation}
{\mathcal F}^\mu(F,n,\nabla n,x,T) =  {\mathcal F}^\mu_e(F,n,T) + {\mathcal F}^\mu_{ni}(n,x) + {\mathcal F}_T(T)+{\mathcal F}^\mu_{F}((\nabla n)F^{-T})
\end{equation}
where ${\mathcal F}_T(T) = {\mathcal F}^\mu_{LC}(Q(T),T)$ and $r(T)=r(Q(T))$.  

According to the neo-classical theory of Bladon, Warner and Terentjev \cite{bladon_1993} that assumes Gaussian statistics of the polymer chain, the entropic elasticity is
\begin{equation} \label{eq:micro}
{\mathcal F}^\mu_e(F,n,T) = \frac{\mu(T) r^{-1/3}}{2} \left(\text{tr } F^T \ell^{-1}(n,r(T))F - 3 \right).
\end{equation}
where $\ell$ is given by (\ref{eq:sl}). Generalizations of this energy have been proposed by Agostoniani and DeSimone \cite{agostiniani_2012} (Also Lee and Bhattacharya \cite{lee_2022}) where they replace the neo-Hookean form with a generalized Mooney-Rivlin form to account for the stiffening at large stretches.  Note that $F^T \ell^{-1}F$ describes the metric of distorsion of the matrix relative to the spontaneous stretch described by $\ell$.  The second term in the free energy, the non-ideality, is a quadratic function of the deviation of the pull-back of the director, $F^{-T}n/|F^{-T}n|$, from a preferred random director field $n_0$ \cite{biggins_2008}.  Finally, the Frank elasticity is quadratic in the spatial gradient of the director ($\nabla$ is the reference gradient operator and hence $(\nabla n) F^{-T}$ is the spatial gradient.).

The elastic term is not quasi-convex and this leads to domain patterns influenced by the disorder in the non-ideal term.  The length-scale of the domain patterns is determined by the competition between the elastic and Frank terms.

\paragraph{Domain pattern scale}
We need to coarse-grain the energy (\ref{eq:micro}) over a domain pattern to obtain the free energy on the macroscale.  It is not possible to compute this explicitly, though one can use bounds \cite{biggins_2009,biggins_2012} and numerical simulations \cite{zhou_2021} as described above (also \cite{barnes_2023} in compression).  We draw on these calculations and postulate the free energy density to be
\begin{equation} \label{eq:meso}
{\mathcal F}(F,P,\Lambda, \Delta) =  {\mathcal F}_e(F,P,\Lambda, \Delta,T) +  {\mathcal F}_r(\Lambda,\Delta,T) + {\mathcal F}_T(T).
\end{equation}
where $F$ is the average deformation gradient  of the domain pattern relative to the isotropic reference state, $P$ is the orientation of the spontaneous stretch and $\Lambda$ and $\Delta$ are the descriptors of the spontaneous stretch.  The first term is the entropic energy of elastomer network, the second the residual energy as a result of non-ideality and any incompatibility in the domains, and the final term is the thermal energy.

Following the neo-classical theory, we take the entropic energy of the elastomer network to depend on the distortion from the spontaneous deformation:\begin{equation}
{\mathcal F}_e(F,P,\Lambda, \Delta) = \tilde {\mathcal F}_e(F^T G^{-1}(P,\Lambda, \Delta) F).
\end{equation}
Here, and below, we suppress temperature from the notation.
This energy is frame-indifferent as required since $G$ is a metric and it transforms as $G \mapsto R^T G R$ under a change of current frame $y \mapsto Ry + c$.
We expect the elastomer network to be isotropic and incompressible, and therefore we expect $\tilde {\mathcal F}$ to depend only on two eigenvalues, or equivalently first two invariants:
\begin{equation}
{\mathcal F}_e(F,P,\Lambda, \Delta) = \tilde {\mathcal F}_e(F^T G^{-1} F) =  \varphi( I_1 (F^T G^{-1} F), I_2 (F^T G^{-1} F) )
\end{equation}
where $I_1(C) = \text{tr}(C)$ and $I_2(C) = \text{tr}( \text{cof }C)$.   It is easy to verify $I_1(C) \ge 3, I_2 (C) \ge 3$ when $\text{det } C=1$.  Therefore, $\varphi$ is only defined on the domain $\{ (I_1, I_2): I_1 \ge 3, I_2 \ge 3\}$.  Further, we assume that $\varphi$ is non-negative, convex in $(I_1, I_2)$ and increasing function of both $I_1$ and $I_2$.  This ensures that an elastic energy density based on $\varphi$ is polyconvex.  A convenient choice, following Agostoniani and DeSimone \cite{agostiniani_2012} (Also Lee and Bhattacharya \cite{lee_2022}) is the generalized Mooney-Rivlin energy
\begin{equation}
\varphi( I_1 , I_2 )= \sum_{i=1}^I C_i ( I_1^{n_i} - 3^{n_i}) +  \sum_{j=1}^J D_j (I_2^{m_j} - 3^{m_j})
\end{equation}
with $C_i, D_j \ge 0$, $n_i, m_j \ge 1$.  The neo-Hookean energy, on with the neo-classical theory is based, is a special case with $I=1, J=0, n_1=1$.

We conclude this section with two results.

\paragraph{Relaxation}
The first result shows that minimizing the energy with respect to the orientation $P$ and descriptors $\Lambda, \Delta$ of spontaneous stretch leads to the relaxation of the Agostoniani and DeSimone \cite{agostiniani_2012} energy.  Specifically, we show that 
\begin{equation}
\overline W (F) : = \min_{\substack{P \  \text{rotation} \\ (\Lambda, \Delta) \in {\mathcal A}_G }} \ \ {\mathcal F}_e(F,P,\Lambda, \Delta) = W^{AD}(F)
\end{equation}
where 
\begin{equation}
W^{AD}(F) = 
\begin{cases}
0 & (\lambda, \delta) \in {\mathcal A}_G\\
\displaystyle{\sum_{i=1}^I} c_i \left( \left( I_1^P \right)^{n_i} - 3 \right) +  
\sum_{j=1}^J d_j \left (\left( I_2^P \right)^{m_j} - 3 \right)  & (\lambda, \delta) \in {\mathcal A}_P 
\\
\displaystyle{\sum_{i=1}^I} c_i \left( \left( I_1^M \right)^{n_i} - 3 \right) +  
\sum_{j=1}^J d_j \left (\left(I_2^M \right)^{m_j} - 3 \right)  & (\lambda, \delta) \in {\mathcal A}_M 
\end{cases} \ .
\end{equation}
with
\begin{eqnarray} 
&I_1^P = \displaystyle{ 2 \frac{r^{1/6}}{\delta} + \frac{\delta^2}{r^{1/3}}}, &
I_1^M = \frac{\lambda^2}{r^{2/3}} + \frac{r^{1/3}\delta^2}{\lambda^2}+ \frac{r^{1/3}}{\delta^2}, \\
&I_2^P = \displaystyle{  2 \frac{r^{1/6}}{\delta} + \frac{\delta^2}{r^{1/3}}}, &
I_2^M = \frac{r^{2/3}}{\lambda^2}+ \frac{\lambda^2}{r^{1/3}\delta^2}+ \frac{\delta^2}{r^{1/3}}, \\
&{\mathcal A}_P = \{ (\lambda, \delta): \min\{r^{1/6}, r^{-1/2} \lambda^2\} \le \delta \le \lambda^2 \} ,&
{\mathcal A}_M = \{ (\lambda, \delta): \sqrt{\lambda} \le  \delta <  r^{-1/2} \lambda^2 \},
\end{eqnarray}
and $\lambda = \lambda(F), \delta = \delta(F)$.
It follows that in the neo-classical setting, $I=1, J=0, n_1=1$,
\begin{equation}
\overline W (F) = \min_{\substack{P \  \text{rotation} \\ (\Lambda, \Delta) \in {\mathcal A}_G }} \ \ {\mathcal F}(F,P,\Lambda, \Delta) = W^{DD}(F)
\end{equation}
where $W^{DD}$ is the relaxation computed by Desimone and Dolzmann \cite{desimone_2002} of the neo-classical energy of Bladon, Warner and Terentjev \cite{bladon_1993}.

The first step is to minimize with respect to the orientation $P$ of the spontaneous deformation.  To do so, we recall the polar decomposition $ F = RU$ for a rotation $R$ and positive-definite, symmetric stretch $U$.  Further, we define $\lambda$ to be the principal stretch (largest eigenvalue of $U$) and $\delta$ to be the principal areal stretch (the product of the two largest eigenvalues of $U$).  So, $U = R'U_0(R')^T$ where 
\begin{equation} \label{eq:u0}
U_0 = \begin{pmatrix} \lambda & 0 & 0 \\ 0 & \delta/\lambda & 0 \\ 0 & 0 & 1/\delta \end{pmatrix}
\end{equation}
in a laboratory frame and $R'$ is a rotation.  Therefore, $F =  R R'U_0 (R')^T$ and 
\begin{equation}
F^TG^{-1}F = R'U_0 (R')^T R^T P^T G_0^{-1} P  R R' U_0 (R')^T =   R' U_0 \tilde Q^T G_0^{-1} \tilde Q U_0 (R')^T
\end{equation}
where $\tilde Q := P RR'$.  It follows that 
\begin{equation}
I_1 (F^T G^{-1} F) = \text{tr}(U_0 \tilde Q^T G_0^{-1} \tilde Q U_0), \quad 
I_2 (F^T G^{-1} F) = \text{tr}(U_0^{-1} \tilde Q^T G_0 \tilde Q U_0^{-1}).
\end{equation}

Now, given positive-definite symmetric matrices
\begin{equation}
A = \begin{pmatrix} a_1 & 0 & 0\\ 0 & a_2 & 0 \\ 0 & 0& a_3 \end{pmatrix}, \quad
B = \begin{pmatrix} b_1 & 0 & 0\\ 0 & b_2 & 0 \\ 0 & 0& b_3 \end{pmatrix}
\end{equation}
with $a_1 \ge a_2 \ge a3$ and $b_1 \ge b_2 \ge b_3$,
\begin{equation}
\min_{ R \text{ rotation}} \  \text{tr}(A  R^T B^{-1}  R A) = \frac{a_1^2}{b_1} + \frac{a_2^2}{b_2} + \frac{a_3^2}{b_3}, 
\end{equation}
It follows that 
\begin{eqnarray}
\bar{I}_1 := \min_{P \text{ rotation}} I_1(F^T G^{-1} F) &=&\frac{\lambda^2}{\Lambda^2} + \frac{\delta^2 \Lambda^2}{\lambda^2\Delta^2} + \frac{\Delta^2}{\delta^2}, \\
\bar{I}_2 := \min_{P \text{ rotation}} I_2(F^T G^{-1} F) &=& \frac{\Lambda^2}{\lambda^2} + \frac{\lambda^2\Delta^2}{\delta^2 \Lambda^2}+ \frac{\delta^2}{\Delta^2},
\end{eqnarray}
and the minimum is attained by $P=(RR')^T$.  Finally,  $\varphi$ is an increasing functions of $I_1, I_2$ and therefore,
\begin{eqnarray}
W(F, \Lambda, \Delta) := \min_{P \text{ rotation}} \ {\mathcal F}(F, P, \Lambda, \Delta) &=& \min_{P \text{ rotation}} \ \varphi( I_1 (F^T G^{-1} F), I_2 (F^T G^{-1} F) ) \\
&=& \varphi \left(\frac{\lambda^2}{\Lambda^2} + \frac{\delta^2 \Lambda^2}{\lambda^2\Delta^2} + \frac{\Delta^2}{\delta^2},
\frac{\Lambda^2}{\lambda^2} + \frac{\lambda^2\Delta^2}{\delta^2 \Lambda^2}+ \frac{\delta^2}{\Delta^2} \right) \\
&=& \varphi (\bar{I}_1, \bar{I}_2),
\end{eqnarray}
and the minimum is attained by $P=(RR')^T$.

It remains to show that 
\begin{equation} \label{eq:wmin}
\overline{W}(F) = \min_{(\Lambda, \Delta) \in {\mathcal A}_G } \ W(F, \Lambda, \Delta)
= W^{AD}(F)
\end{equation}
We have the following exhaustive cases:
\begin{enumerate}

\item { Interior of ${\mathcal A}_G$.}  Suppose the minimum in (\ref{eq:wmin}) is attained in the interior of ${\mathcal A}_G$.  Since $\varphi$ is non-negative, increasing and convex, this implies that 
\begin{equation}
\frac{\partial \bar{I}_1}{\partial \Lambda} = \frac{\partial \bar{I}_2}{\partial \Lambda} = \frac{\partial \bar{I}_1}{\partial \Delta} = \frac{\partial \bar{I}_2}{\partial \Delta} = 0.
\end{equation}
The first and third of this imply that $\lambda^4/\Lambda^4 = \delta^2/\Delta^2$ and $\lambda^2/\Lambda^2 = \delta^4/\Delta^4$ which in term imply that $\Lambda=\lambda, \Delta= \delta$.  This in turn implies that $\bar{I}_1 = \bar{I}_2 = 3$ and therefore $W = 0$.  Further, $(\lambda, \delta) \in {\mathcal A}_G$ and therefore $W^{AD} = 0$.  Thus, $W=W^{AD}$ in the interior of ${\mathcal A}_G$.

\item { Boundary with $\Delta = r^{1/6}, r^{1/12} < \Lambda < r^1/6 $.} Suppose the minimum in (\ref{eq:wmin}) is attained on that portion of the boundary where $\Delta = r^{1/6}, r^{1/12} < \Lambda < r^1/6 $.  We  have
\begin{equation}
\frac{\partial \bar{I}_1}{\partial \Lambda} = \frac{\partial \bar{I}_2}{\partial \Lambda} = 0, 
\ \frac{\partial \bar{I}_1}{\partial \Delta} \le 0, \ \frac{\partial \bar{I}_2}{\partial \Delta} \le 0.
\end{equation}
The first implies $\lambda^4 = \Lambda^4 \delta^2/r^{1/3}$ which in turn implies that $\bar{I}_1 = I_1^P, \bar{I}_2 = I_2^P$ and so $\overline{W} = \varphi(I_1^P, I_2^P)$.  Further, the restriction on $\Lambda$ implies that $\delta \le \lambda^2 \le r^{1/2} \delta$.  Finally, the third inequality above implies $\delta^3 \ge r^{2/3}\lambda^2 \Lambda^2 = r^{1/2} \Lambda^2 \ge r^{1/6}$.  Putting all of this together, we conclude that $(\lambda, \delta) \in {\mathcal A}_P$ and $\overline{W} = W^{AD}$ on ${\mathcal A}_P$.

\item { Corner $\Delta = r^{1/6}, \Lambda = r^{1/3}$. }  Suppose the minimum in (\ref{eq:wmin}) is attained at $\Delta = r^{1/6}, \Lambda = r^{1/3}$.  It follows $\bar{I}_1 = I_1^M, \bar{I}_2^M$ so that $\bar{W} = \varphi( I_1^M, I_2^M)$.  Further, we have
\begin{equation}
\frac{\partial \bar{I}_1}{\partial \Lambda} \le 0, \ \frac{\partial \bar{I}_2}{\partial \Lambda} \le 0, \ 
\frac{\partial \bar{I}_1}{\partial \Delta} \le 0, \ \frac{\partial \bar{I}_2}{\partial \Delta} \le 0.
\end{equation}
Together, the first and third implies that $(\lambda, \delta) \in {\mathcal A}_M$.  Thus, $\overline{W} = W^{AD}$ on ${\mathcal A}_M$.

\item { Corner $\Delta = r^{1/6}, \Lambda = r^{1/12}$. }
Suppose the minimum in (\ref{eq:wmin}) is attained at $\Delta = r^{1/6}, \Lambda = r^{1/12}$.  Arguing as above, $\delta = r^{1/6}, \lambda = r^{1/12}$ so that $(\lambda,\delta) \in {\mathcal A}_G$ and $\overline{W} = 0 = W^{AD}$.

\item { Boundary with $\Delta < r^{1/6}$}.  Suppose the minimum in (\ref{eq:wmin}) is attained on those portions of the boundary of 
${\mathcal A}_G$ where $\Delta < r^{1/6}$, i.e., where $\Lambda = \Delta^2$ or $\Lambda = \sqrt{\Delta}$.  Similar arguments lead to the conclusion that $(\lambda,\delta) \in {\mathcal A}_G$ and $\overline{W} = 0 = W^{AD}$.
\end{enumerate}
This completes the proof that $\overline{W} = W^{AD}$.

\paragraph{Stress}
Finally, we show that minimization with respect the orientation $P$ and interior minimization with respect to the descriptor $\Lambda$ of microstructure leads to an equality of the first two principal stresses.  Recall that in an isotropic, incompressible hyper-elastic body with energy density $W$, we may write the principal components of the Cauchy stress to be
\begin{equation} \label{eq:si}
\sigma_i = \lambda_i \frac{dW}{d\lambda_i} - p, \quad \quad i = 1, 2, 3.
\end{equation}
Now, as shown earlier,
\begin{equation}
\bar {\mathcal F}(F, \Lambda, \Delta) := \min_{P \text{ rotation}} \ {\mathcal F}(F,P,\Lambda,\Delta) = \varphi \left(\frac{\lambda^2}{\Lambda^2} + \frac{\delta^2 \Lambda^2}{\lambda^2\Delta^2} + \frac{\Delta^2}{\delta^2},
\frac{\Lambda^2}{\lambda^2} + \frac{\lambda^2\Delta^2}{\delta^2 \Lambda^2}+ \frac{\delta^2}{\Delta^2} \right).
\end{equation}
It follows that the first two principal stresses are given by 
\begin{equation} \label{eq:stress}
\sigma_1 = 2 \left ( \varphi_{,1} \frac{\lambda^2}{\Lambda^2} - \varphi_{,2} \frac{\Lambda^2}{\lambda^2} \right) - p, \quad \quad
\sigma_2 = 2 \left ( \varphi_{,1} \frac{\delta^2 \Lambda^2}{\lambda^2\Delta^2} - \varphi_{,2}\frac{\lambda^2\Delta^2}{\delta^2 \Lambda^2} \right) - p
\end{equation}
where $\varphi_{,1}$ and $\varphi_{,2}$ describe the partial derivatives with respect to $I_1$ and $I_2$ respectively.  Further, equilibrium with respect to $\Lambda$ requires that
\begin{eqnarray}
0 &=& \frac{\partial \bar {\mathcal F}}{\partial \Lambda} = 
\varphi_{,1} \left( -2 \frac{\lambda^2}{\Lambda^3} + 2 \frac{\delta^2 \Lambda}{\lambda^2\Delta^2} \right) 
+ \varphi_{,2} \left( 2 \frac{\Lambda}{\lambda}^2 - \frac{\lambda^2 \Delta^2}{\delta^2 \Lambda^3} \right)\\
&=& \frac{1}{\Lambda} \left(
2 \left ( \varphi_{,1} \frac{\lambda^2}{\Lambda^2} - \varphi_{,2} \frac{\Lambda^2}{\lambda^2} \right)
- 2 \left ( \varphi_{,1} \frac{\delta^2 \Lambda^2}{\lambda^2\Delta^2} - \varphi_{,2}\frac{\lambda^2\Delta^2}{\delta^2 \Lambda^2} \right) \right)\\
&=&  \frac{1}{\Lambda}  (\sigma_1 - \sigma_2).
\end{eqnarray}
The result follows.

\subsection{Domain pattern evolution} \label{sec:evol}

The orientation of the director at the microscale, and consequently the domain pattern at the mesoscale, are not fixed, but can evolve in response to applied loads in an I-PLCE.  Tokumoto {\it et al.} \cite{tokumoto_2021} conducted extensive bi-axial mechanical tests on thin I-PLCE sheets while monitoring the evolution of the microstructure using wide-angle x-ray scattering (WAXS).  Since the I-PLCE is isotropic and incompressible, a general state of deformation is characterized by two principal stretches.  Thus, a biaxial test where the two stretches are prescribed independently -- as implemented in Tokumoto {\it et al.} -- suffices to describe arbitrary deformations.  They observed an intriguing in-plane liquid-like behavior where the film did not develop shear stress even when a shear strain is imposed.  Specifically, the true (Cauchy) stress in the two directions of principal stretch remain equal even when the imposed principal stretches are different.  Further, the value of this (equal) true stress depends only on the product of the two largest principal stretches and is independent of the largest principal stretch.  In other words, $\Lambda$ evolves to negate the shear stress but  $\Delta$ determines the  (equal) true stress.  The WAXS observations provided further evidence that the domain patterns re-arranged in the plane to negate the applied difference in principal stretches.  

\begin{figure}
\centering
\includegraphics[width=5in]{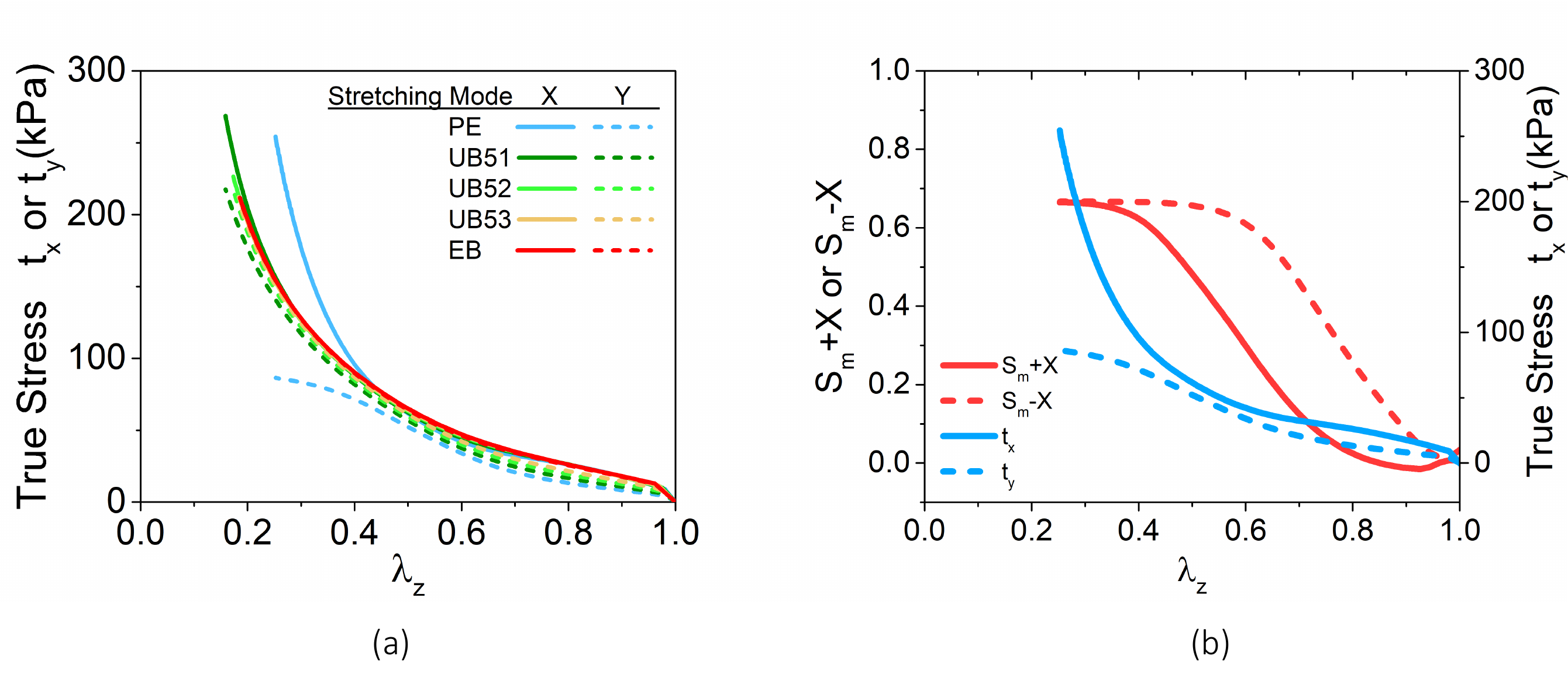}
\caption{Domain pattern evolution.  Adapted from Figure 8 of Tokumoto {\it et al.} \cite{tokumoto_2021} \label{fig:evol}}
\end{figure}

Tokumoto {\it et al.} \cite{tokumoto_2021} also reports the results of detailed numerical simulations of the domain patterns in an I-PLCE using a model and numerical method detailed in Zhou and Bhattacharya \cite{zhou_2021}.  Figure \ref{fig:evol} (adapted from Figure 8 of Tokumoto {\it et al.} \cite{tokumoto_2021}) shows some details.   Figure \ref{fig:evol}(a) shows that the two in-plane true stresses are equal and determined only by the product of the two principal stretches ($\delta = \lambda_x \lambda_y = 1/ \lambda_z$). Figure \ref{fig:evol}(b) focusses on planar extension (PE).   It shows that both $S_m -X$ and $S_m + X$ evolve initially (approximately $\lambda_z = 1/\Delta \ge 0.7$).  The two principal stresses are small, but also differ: this difference is small because they are both small.  There is then an intermediate regime (approximately $0.7 \ge \lambda_z = 1/\Delta \ge 0.4$) where $S_m -X$ is constant but $S_m +X$ continues to rise.  The principal stresses continue to rise but are equal to each other as they do so.  Finally, (approximately $\lambda_z = 1/\Delta \le 0.4$), both $S_m -X$ and $S_m + X$ saturate and remain constant.  The two stresses diverge.  We conclude that the level of the principal stresses is governed by $S_m + X$ and consequently $\Delta$ while the difference depends on $S_m - X$.  In other words, $S_m, X$ evolve as necessary to accommodate the imposed deformation with zero shear stress.  Recalling the identification of $\Lambda$ with $S_m$ and $\Delta$ with $S_m, X$, we 
conclude that $\Lambda$ evolves rapidly to equalize the true stresses while $\Delta$ evolves slowly and controls the true stress. 

\section{Constitutive model} \label{sec:model}

In this section, we propose, demonstrate and validate a constitutive model for isotropic-genesis polydomain liquid crystal elastomer (I-PLCE) that describes the unique properties of the material at the engineering or macro scale.  The model incorporates the microscale and macroscale physics described in the previous section by introducing state variables.  These variables and their evolution describes the implications of the mesoscale and microscale dynamics in a coarse-grained manner at the engineering scale.

In this section, we limit ourselves to isothermal processes and therefore incorporate temperature through the material parameters.   \ref{sec:dissipation} describes the generalization to other processes, and Section \ref{sec:conc} describes some of the open challenges.

\subsection{Formulation}

Consider a macroscale specimen of I-PLCE, and consider the stress-free isotropic state as the natural reference configuration.  The state of the specimen in the current configuration  is characterized by a deformation gradient $F$ relative to the  reference configuration and two state variables $\Lambda$ and $\Delta$\footnote{We introduce the state variables empirically here, but are motivated by the multiscale framework of the Section \ref{sec:back}.  Specifically, they were introduced as descriptors of spontaneous stretch in Section \ref{sec:sd} and are related to the underlying domain pattern.}.  The material is incompressible and hence $\text{det} \ F = 1$.   Further, we restrict the state variables to the ${\mathcal A}_G$ defined in (\ref{eq:ag}).

We postulate a free energy density (per unit reference volume)
\begin{eqnarray} \label{eq:w}
W(F, \Lambda, \Delta) &=&  W_e(F,\Lambda,\Delta) + W_r(\Lambda,\Delta)
\end{eqnarray}
where $W_e$ is the elastic energy stored in the polymer network as a result of deformation relative to the spontaneous deformation of the domains, and $W_r$ is the residual energy during the spontaneous deformation as a result of any incompatibility in the domains.
We assume that 
\begin{eqnarray}
 W_e(F,\Lambda,\Delta) &=& 
\varphi \left(I_1(F^TGF) , I_2(F^TGF) \right)\\
&=& \varphi \left(\frac{\lambda^2}{\Lambda^2} + \frac{\delta^2 \Lambda^2}{\lambda^2\Delta^2} + \frac{\Delta^2}{\delta^2},
\frac{\Lambda^2}{\lambda^2} + \frac{\lambda^2\Delta^2}{\delta^2 \Lambda^2}+ \frac{\delta^2}{\Delta^2} \right).
\end{eqnarray}
where $\varphi$ is non-negative, convex and increasing function of its arguments,  $I_1, I_2$ are the first two principal invariants, $G = R^TQT G_0 QR$,  for rotations $Q, R$ determined as $F=QRU_0 R^T$ with $G_0$ and $U_0$ as in (\ref{eq:g0}) and (\ref{eq:u0}) respectively, and $\lambda = \lambda(F)$ is the principal stretch and $\delta = \delta(F)$ is the principal areal stretch (product of the two largest eigenvalues) associated with $F$\footnote{Note that we have assumed that the overall reorientation $P$ of the domain pattern evolves rapidly and therefore can be minimized out of the problem.}.  We further assume that 
\begin{eqnarray} \label{eq:wr}
 W_r(\Lambda,\Delta) = C \frac{\Delta - 1}{\left(r^{1/6}-\Delta\right)^k}.
\end{eqnarray}
where $C>0, k>1$ are constitutive constants to be determined.   This term is small near the isotropic state ($\Delta =1$), but increases as the the domain patterns evolves away from it and blows up as the domain patterns become planar  ($\Delta = r^{1/6}$).   Experiment and simulations \cite{tokumoto_2021,zhou_2021} show that the randomness in the cross-linking of the I-PLCE make the isotropic state ($\Delta =1$) the  ground state.  Further, there are many metastable states close to this state and therefore it is easy to change the domain patterns near the isotropic state.  This becomes harder as the domain patterns move away from the isotropic state, and it requires larger loads as the domain pattern approaches a planar arrangement ($\Delta = r^{1/6}$).  This motivates this particular form for $W_r$.

We postulate that the stress is the sum of a conservative (elastic) contribution associated with $W$, and a dissipative contribution.  In this work, we take the latter to be linear or Newtonian.  So, the Cauchy stress
\begin{equation} \label{eq:cauchy}
\sigma = -p I + \frac{\partial W}{\partial F} F^T + \nu D
\end{equation}
where $p$ is an unknown hydrostatic pressure (due to incompressibility), $\nu$ is the viscosity and $D = \text{symm}(\dot FF^{-1})$ is the rate of deformation.  The elastic contribution can be calculated easily in the principal basis through (\ref{eq:si}).  We assume that the stress satisfies the equation of equilibrium
\begin{equation}
\text{div }\sigma + b = 0
\end{equation}
subject to boundary conditions where $b$ is the body force per unit reference volume.

It remains to specify the evolution of the state variables.  We identify the energetic or thermodynamic driving forces associated with $\Lambda$ and $\Delta$ to be 
\begin{equation} \label{eq:state1}
d_\Lambda = - \frac{\partial W}{\partial \Lambda}, \quad \quad
d_\Delta = - \frac{\partial W}{\partial \Delta}
\end{equation} respectively (see \ref{sec:dissipation} for details).  We postulate that the rate of change of these variables depends on their respective driving foces:
\begin{eqnarray} \label{eq:state2}
\dot \Lambda = K_\Lambda (d_\Lambda), \quad \quad
\dot \Delta = K_\Delta (d_\Delta)
\end{eqnarray} 
where $k_\Lambda, k_\Delta$ are functions that satisfy $x K_\Lambda(x) \ge 0, x K_\Delta(x) \ge 0 \ \forall x$ to satisfy the dissipation inequality.  In this work, we take these functions to be linear, and so
\begin{eqnarray} \label{eq:state3}
\nu_\Lambda \dot \Lambda = - \frac{\partial W}{\partial \Lambda}, \quad \quad
\nu_\Delta \dot \Delta = -  \frac{\partial W}{\partial \Delta}
\end{eqnarray} 
where $\nu_\Lambda \ge 0, \nu_\Delta \ge 0$.  Recalling the discussion in Section \ref{sec:evol}, we assume that $\nu_\Lambda << \nu_\Delta$.  See \ref{sec:dissipation} for a more general forms.

\subsection{Demonstration and validation} \label{sec:valid}
\begin{table}
\centering
\begin{tabular}{lc}
\hline
\multicolumn{2}{c}{Parameters used for demonstration and validation}\\
\hline
Shear modulus $C_1, \mu$ & $4.94 \times 10^4$ Pa\\
LCE anisotropy parameter, $r$ & 9.14\\
Viscosity, $\nu$ & 0\\
Hardening coefficient, $C$ & 298 Pa\\
Hardening exponent $k$ & 2\\
Kinetic coefficient, $\nu_\Delta$ & $2.18 \times 10^7$ Pa\\
Kinetic coefficient, $\nu_\Lambda = \nu_\Delta/100 $ \hspace{0.5in} & $ 2.18\times 10^5$ Pa\\ \\
\hline
\multicolumn{2}{c}{Additional parameters used in the \texttt{ABAQUS} implementation}\\
\hline
Bulk modulus $\kappa$ & $4.94 \times 10^8$ Pa\\
Viscosity, $\nu$ & $1 \times 10^6 $ Pa $\cdot$ s\\
\hline
\end{tabular}
\caption{Parameters used to demonstrate the model and validate it against the experimental observations in \cite{tokumoto_2021} (Section \ref{sec:valid}) as well as the additional parameters used in the \texttt{ABAQUS} implementation (Section \ref{sec:verif}).\label{tab:param}}
\end{table}

We now demonstrate the model using biaxial stretch, and validate it against  the experimental observations in \cite{tokumoto_2021} using biaxial stretch experiments.  We use the neo-classical version of the elastic energy so that $I=1, J=0, n_1=1$, and we also assume that the viscosity is zero.  We use the parameters shown in Table \ref{tab:param}: these parameters were chosen to fit the experiments using a gradient descent method.  The model was implemented in \texttt{MATLAB} \cite{matlab}.  The evolution equation is treated explicitly.  

\begin{figure}
\centering
\includegraphics[width=6.25in]{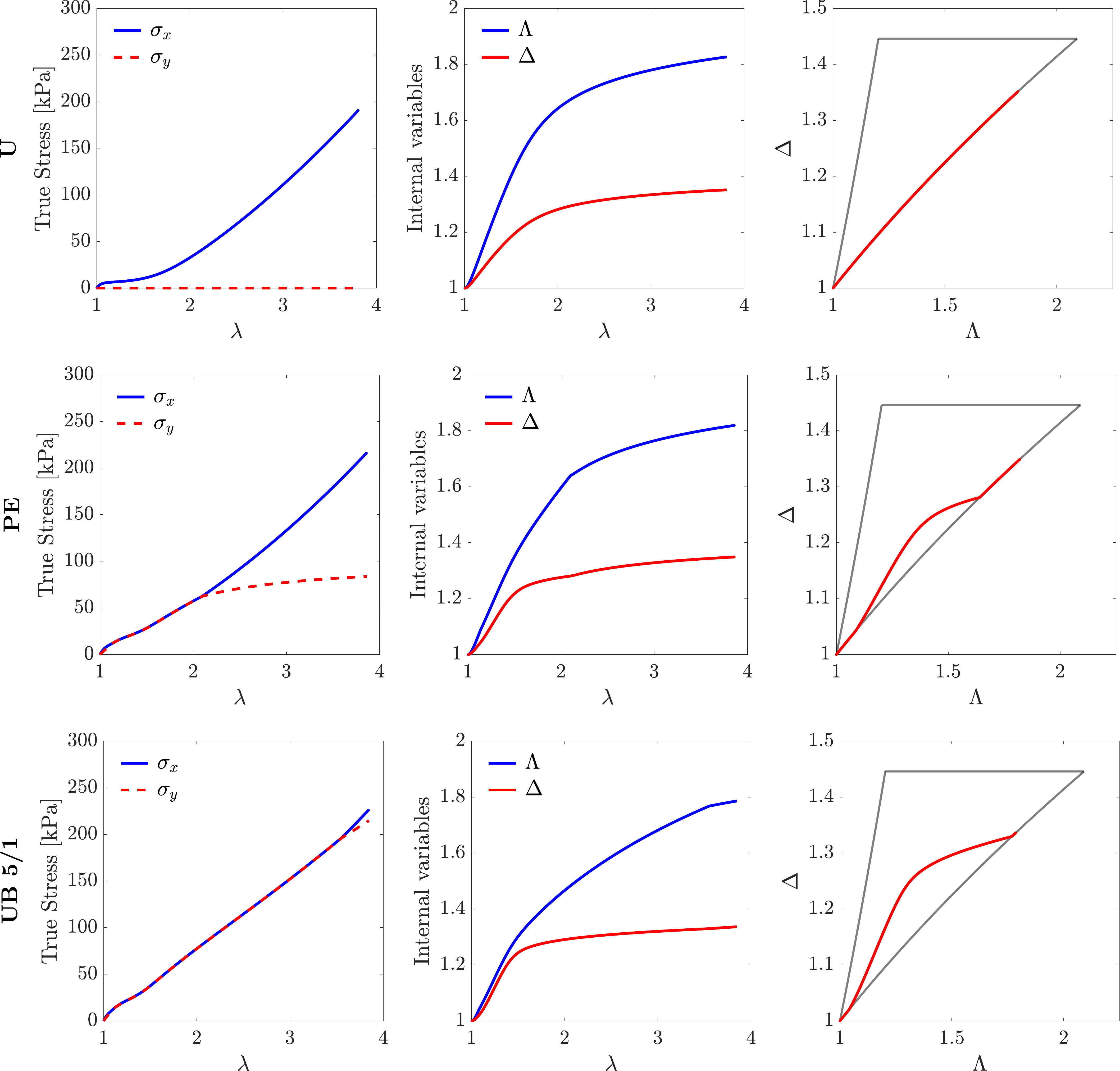}
\caption{Demonstration of the proposed model.  Each row represents a different loading mode described in Table \ref{tab:lm}.  The first column shows the true stress vs.\ stretch, the second column the state variables vs. stretch and the final column the evolution of the state variables.  The figure continues below \label{fig:moddem}}
\end{figure}

Figure \ref{fig:moddem} shows the results of various uniaxial and biaxial stretch tests at a loading rate of $10^{-3}$ s$^{-1}$.   In the uniaxial stress test (U), the stretch history is prescribed in one direction while the lateral directions are assumed to be traction-free.  In the biaxial stretch tests (PE, UB X/Y, EB), the stretch is prescribed along two perpendicular axes, and the third direction is assumed to be stress-free.  Further details are provided in Table \ref{tab:lm}.   Figure \ref{fig:moddem} shows the true (Cauchy) stress components vs. stretch, the state variables vs. stretch and the state variable trajectory for each case.

\begin{table}
\centering
\begin{tabular}{ccc}
\hline
label & loading mode & principal stretch \\
\hline
U & uniaxial stress & $\lambda_x = \lambda$\\
PE & biaxial stretch & $\lambda_x = \lambda, \lambda_y = 1$\\
UB X/Y & biaxial stretch & $\lambda_x = \lambda, \lambda_x - 1/(\lambda_y -1) = X/Y$\\
EB & biaxial stretch & $\lambda_y=\lambda_x = \lambda$\\
\hline
\end{tabular}
\caption{Loading modes used for demonstration and validation \label{tab:lm}}
\end{table}

\renewcommand{\thefigure}{\arabic{figure} (cont.)}
\setcounter{figure}{3}

\begin{figure}
\centering
\includegraphics[width=6.25in]{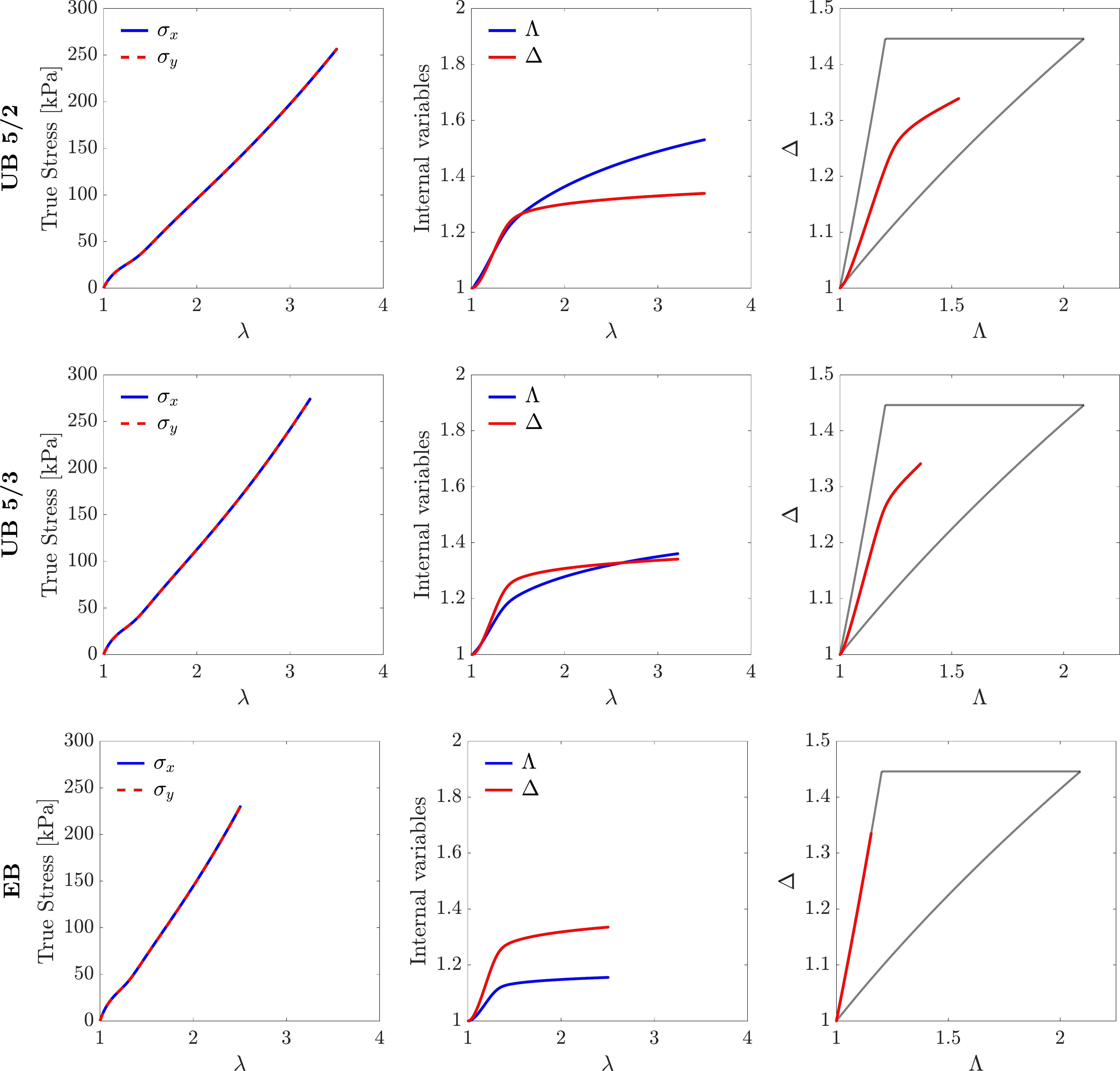}
\caption{Demonstration of the proposed model.  Each row represents a different loading mode described in Table \ref{tab:lm}.  The first column shows the true stress vs.\ stretch, the second column the state variables vs. stretch and the final column the evolution of the state variables.}
\end{figure}
\renewcommand{\thefigure}{\arabic{figure}}

In U, we observe that the uniaxial stress rises initially with stretch, then plateaus as the state variables begin to evolve.  The stress rises again as state variables saturate.  The state variable evolves along the boundary $\Delta = \sqrt{\Lambda}$ of $\mathcal{A}_G$.  In PE, both components of Cauchy stress remain the same even though the stretch components are different.  Further, there is only a very small plateau in the stress.  The state variable $\Delta$ rises and saturates rapidly and the saturation corresponds to the end of the plateau.  The state variable $\Lambda$ continues to rise until it saturates, and this divergence coincides with the divergence of the stress.  The state variables evolve into the interior of $\mathcal{A}_G$ before returning to the boundary.  These trends continue in UB X/Y with the stresses being equal for progressively longer with decreasing ratio X/Y, with the saturation of the state variable $\Lambda$ being progressively delayed and the evolution path drifting towards the other boundary of $\mathcal{A}_G$.  Finally, in EB, the stresses always remain equal and the state variables evolve along the boundary $\Delta = \Lambda^2$ of $\mathcal{A}_G$.

\begin{figure}
\centering
\includegraphics[width=4.5in]{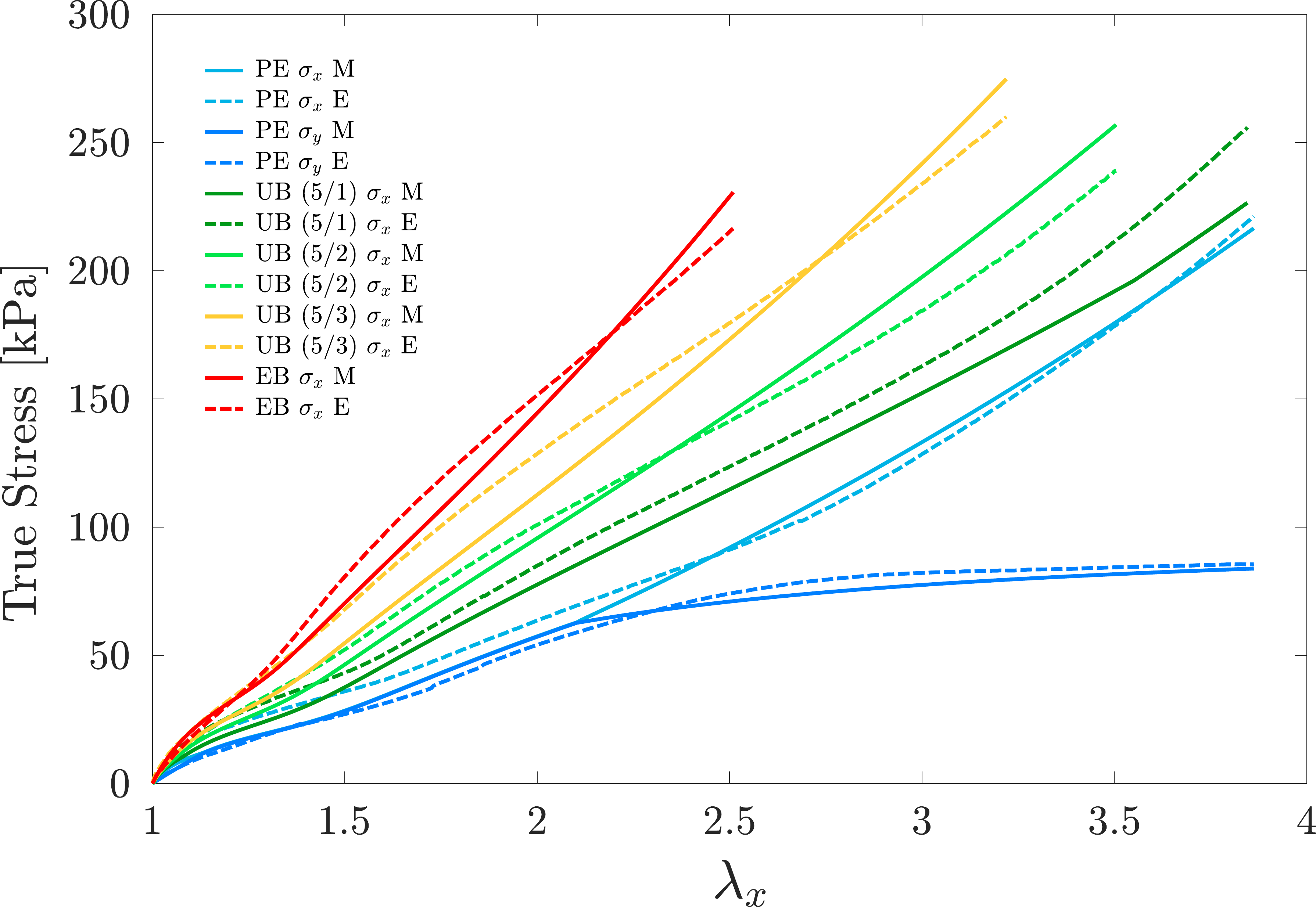}
\caption{Validation \label{fig:valid}}
\end{figure}

We turn to validation in Figure \ref{fig:valid} where we compare the predictions of the model for parameters in Table \ref{tab:param} with the experimental observations in Tokomuto {\it et al.}  \cite{tokumoto_2021}.  We note that the model is able to predict a complex set of observations.  This is particularly striking since the model has so few parameters, and a number of these ($\mu, r$) can be independently estimated from first principles.  We believe that this good approximation reflects the fact that the model implicitly incorporates the underlying physics of microstructure evolution as discussed in Section \ref{sec:back}.

\subsection{Parameter study}

We now demonstrate various aspects of the model by conducting a parameter study.  In each set of tests, we vary one parameter while holding the rest at the values specified in Table \ref{tab:param}.

\begin{figure}
    \centering
    \includegraphics[width=3.0in]{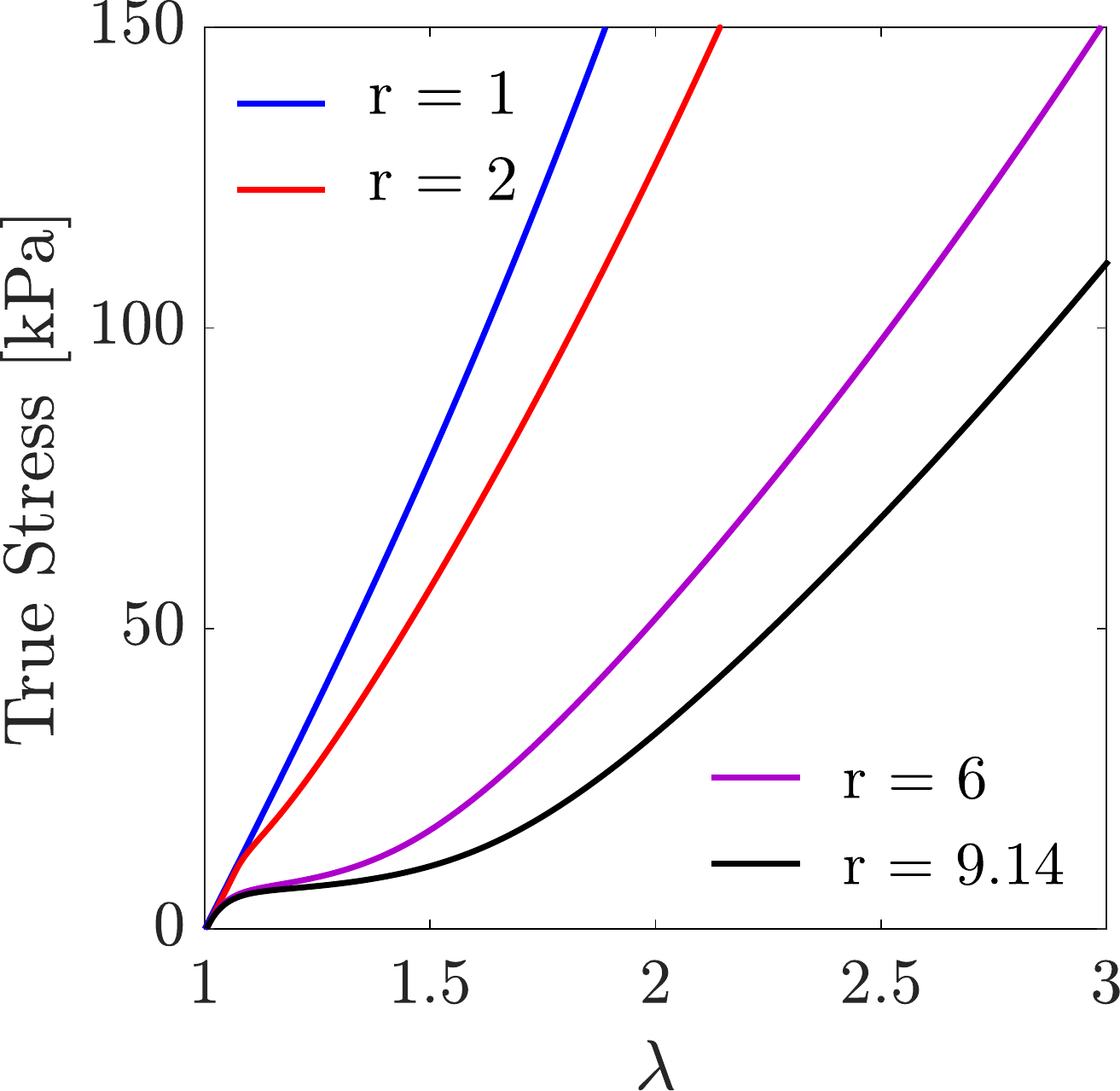}
    \caption{The role of the anisotropy parameter $r$ in U.  This parameter controls the size and level of the plateau in the stress-strain response. }
    \label{fig:r}
\end{figure}

We begin with the anisotropy parameter $r$, and this is shown in Figure \ref{fig:r} for the uniaxial stress U.  We see that this parameter controls the extent and the level of the stress plateau in the stress-stretch response.  $r=1$ corresponds to a neo-Hookean material and we have purely elastic behavior.  As $r$ increases, the plateau appears and grows and the stress level of the plateau decreases.

\begin{figure}
    \centering
    \includegraphics[width=6.25in]{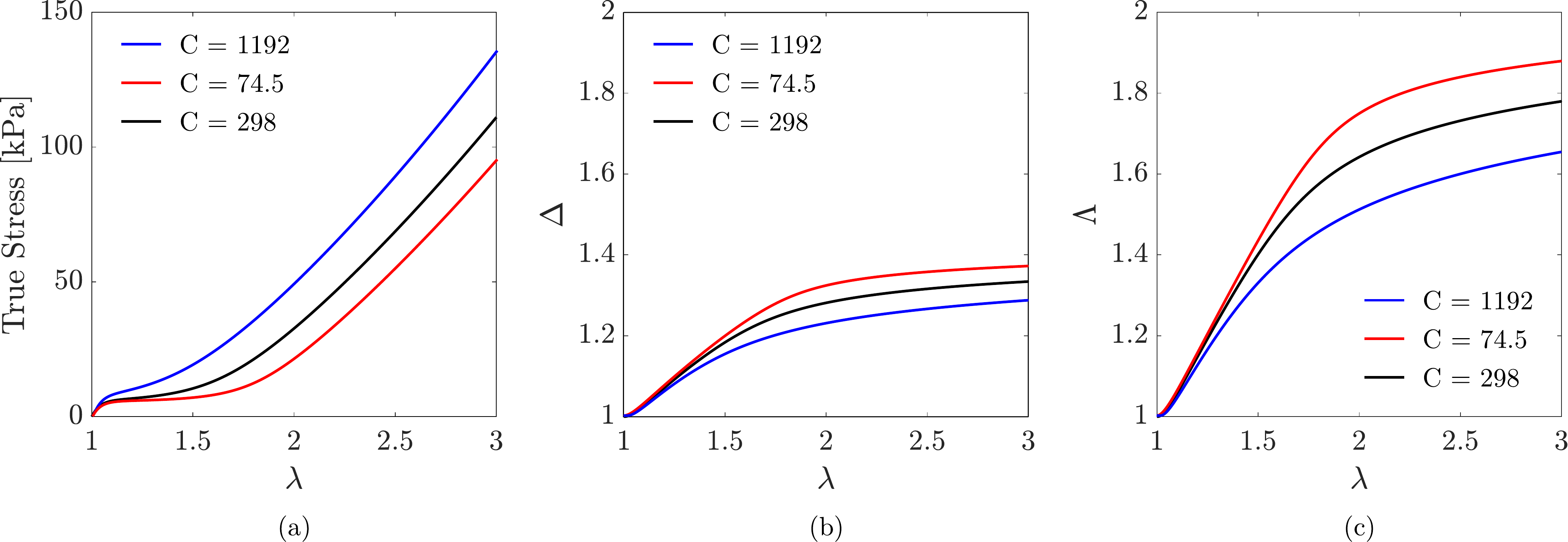}
    \caption{The role of the coefficient $C$ to the residual energy in U.  (a) Stress vs. stretch, (b) State variable $\Delta$ vs. stretch and (c) State variable $\Lambda$ vs. stretch.  Increasing $C$ impedes the evolution of the state variables and therefore stiffens the stress-stretch response.}
    \label{fig:c}
\end{figure}

Figure \ref{fig:c} shows the effect of the coefficient $C$ to the residual energy (\ref{eq:wr}), again on uniaxial stress U. We  see in Figure \ref{fig:c}(a) that  increasing $C$ decreases and raises the plateau in the stress-stretch response  This is expected since increasing $C$ increases the energetic penalty of changing $\Delta$ away from $1$, and this is evident in Figure \ref{fig:c}(b).  This also impedes the evolution of $\Lambda$, Figure \ref{fig:c}(c):  Recall from Figure \ref{fig:moddem} that the state variables follow the line $\Delta = \sqrt{\Lambda}$ in this uniaxial loading and therefore the evolution of $\Lambda$ is linked to that of $\Delta$.

\begin{figure}
    \centering
    \includegraphics[width=6.25in]{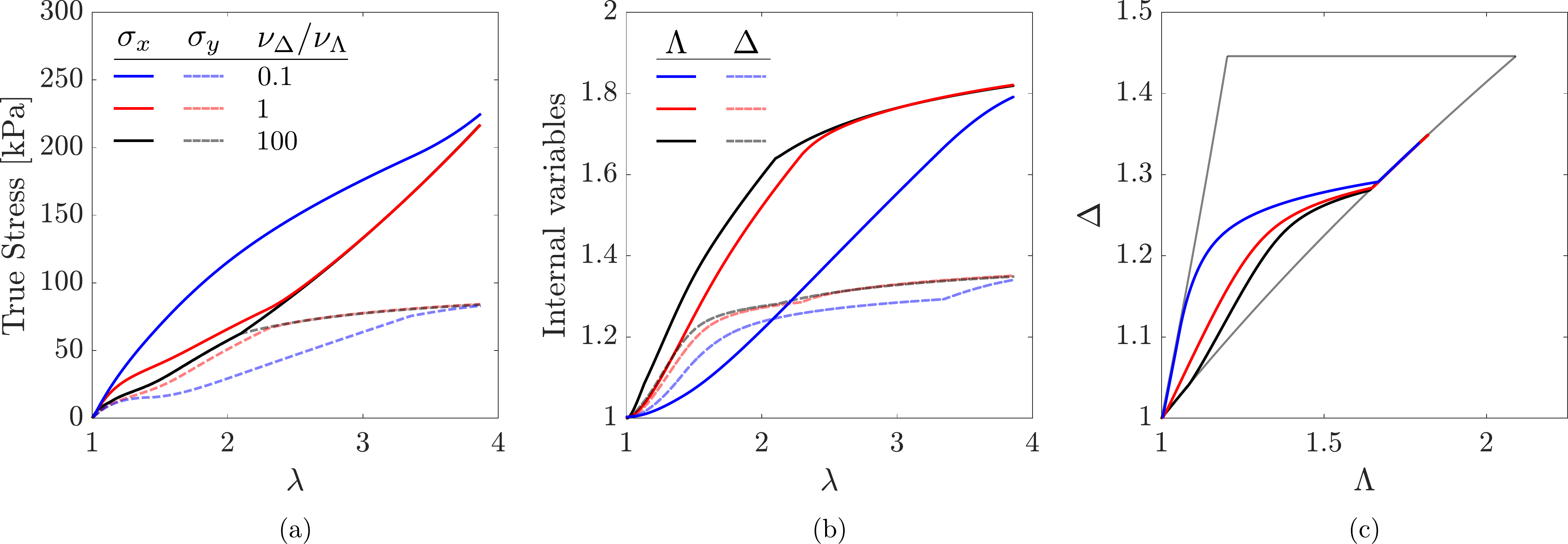}
    \caption{The role of the ratio $\nu_\Delta/\nu_\Lambda$ during PE loading.  (a) Stress vs. stretch, (b) State variable vs. stretch, (c) Evolution of the state variable.  A large ratio ensures the equality of the two components true stress while a small ratio causes their divergence.}
    \label{fig:arat}
\end{figure}

We now turn to the kinetic parameters $\nu_\Lambda$ and $\nu_\Delta$ that control the rate of evolution of the state variables.  Rescaling them by the same factor corresponds to a rescaling.  However, changing their relative magnitude has a profound effect on the material response as shown in Figure \ref{fig:arat} for PE.  In our demonstration, we took $\nu_\Delta/\nu_\Lambda=100$ so that $\Lambda$ would evolve faster than $\Delta$.  This results in the observed in-plane liquid-like behavior where the true stress depends only on the areal stretch and not the individual stretches.  Consequently, the two components of true stress are equal even though the stretches are different.  This is again shown in Figure \ref{fig:arat}.  We see in the figure that the two components of stress diverge as the ratio $\nu_\Delta/\nu_\Lambda$ decreases.

\begin{figure}
    \centering
    \includegraphics[width=3.0in]{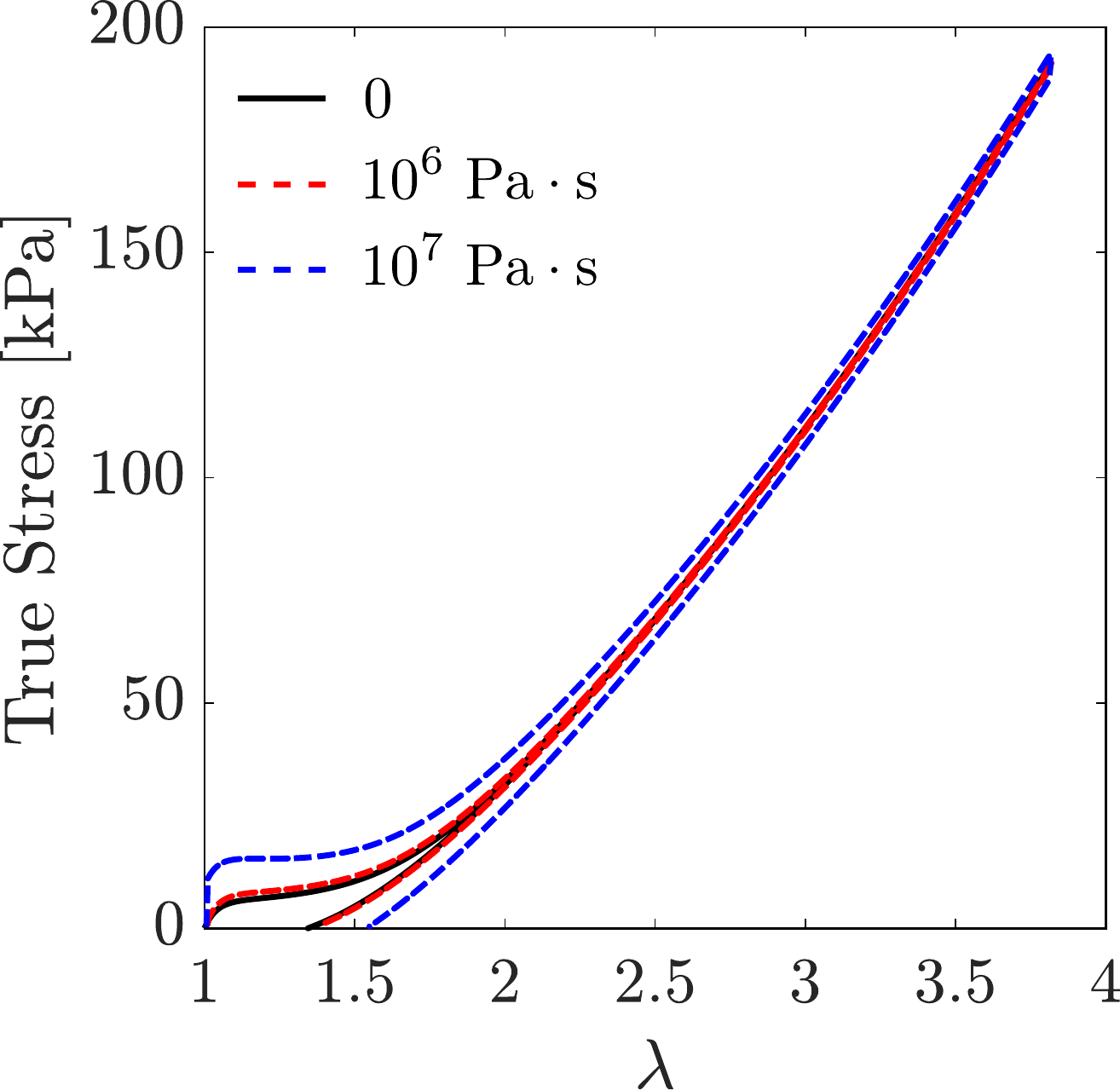}
    \caption{The role of viscosity, $\nu$ during uniaxial loading.}
    \label{fig:visc}
\end{figure}

Finally, the effect of polymer viscosity on the uniaxial stress U is shown in Figure \ref{fig:visc}. This parameter controls the amount of hysteresis in the material response. As expected, we see that as polymer viscosity increases, there is more hysteresis in the stress-stretch response. 

\section{Computational implementation} \label{sec:comp}

Having established the model, we turn to its implementation in the commercial finite element platform \texttt{ABAQUS} \cite{abaqus}.  We first extend the model to be compressible (with a high bulk to shear modulus ratio) for ease of implementation.  We use an implicit approach (\texttt{ABAQUS} Standard), and therefore compute the material Jacobian (through the Jaumann rate of the Kirchhoff stress).  We then verify the implementation and demonstrate it on the simple problem of torsion.  The model has been used to understand experimental observations on torsion-induced instabilities \cite{urayama_2023}, adhesion \cite{maghsoodi_2023} and Hertz contact \cite{terentjev_2023}.

\subsection{Formulation}

We postulate an energy density $W=W_e+Wr$ as in (\ref{eq:w}) with 
\begin{eqnarray}
W_e (F, \Lambda, \Delta) 
&=& {\mu \over 2}\left((\det F)^{-2/3} \ \text{tr }\left( F^T  G^{-1}  F\right) -3 \right)+ {\kappa \over 2}\left(\ln(\det F) \right)^2 \\
&=& {\mu \over 2}\left( (\lambda_1 \lambda_2 \lambda_3)^{-2/3} \left(\frac{\lambda_1^2}{\Lambda^2}+ \frac{\lambda_2^2 \Lambda^2}{\Delta^2}+   \Delta^2 \lambda_3^2 \right)-3 \right)+ {\kappa \over 2}(\ln(\lambda_1 \lambda_2 \lambda_3))^2
\end{eqnarray}
where $\mu$ and $\kappa$ are the shear and bulk modulus respectively, we have taken $G$ as before and $\lambda_1 \ge \lambda_2 \ge \lambda_3$ are the ordered principle values of $F$.  Note that as $\kappa/\mu \to \infty$, this tends to the neo-classical energy density with $C_1=\mu$.  $W_r$ is given by (\ref{eq:wr}) as before.

The Cauchy stress is now
\begin{equation}
\sigma = \frac{1}{\det F} \frac{\partial W}{\partial F} F^T + \nu D
\end{equation}
where $\nu$ is the viscosity and $D$ the deformation rate.

\texttt{ABAQUS} requires the so-called consistent Jacobian $\mathbb{J}$ (called DDSDDE in \texttt{ABAQUS}) defined through the relation
\begin{equation} \label{eq: ddsdde}
\overset{\triangledown}{\tau} = (\det F) {\mathbb J} D \quad (\overset{\triangledown}{\tau}_{ij} = (\det F) {\mathbb J}_{ijkl} D_{kl} \text{ in indicial notation})
\end{equation}
where $\overset{\triangledown}{\tau} = \dot \tau - W \tau + \tau W$ is the Jaumann rate of the Kirchhoff stress ($\tau = (\det F)\sigma$)
and $W = (L- L^T)/2$ is the spin.  An explicit formula for $\mathbb J$ is provided in the \ref{sec:cj}.
 
\subsection{Verification} \label{sec:verif}

\begin{figure}
\centering
\includegraphics[width=5 in]{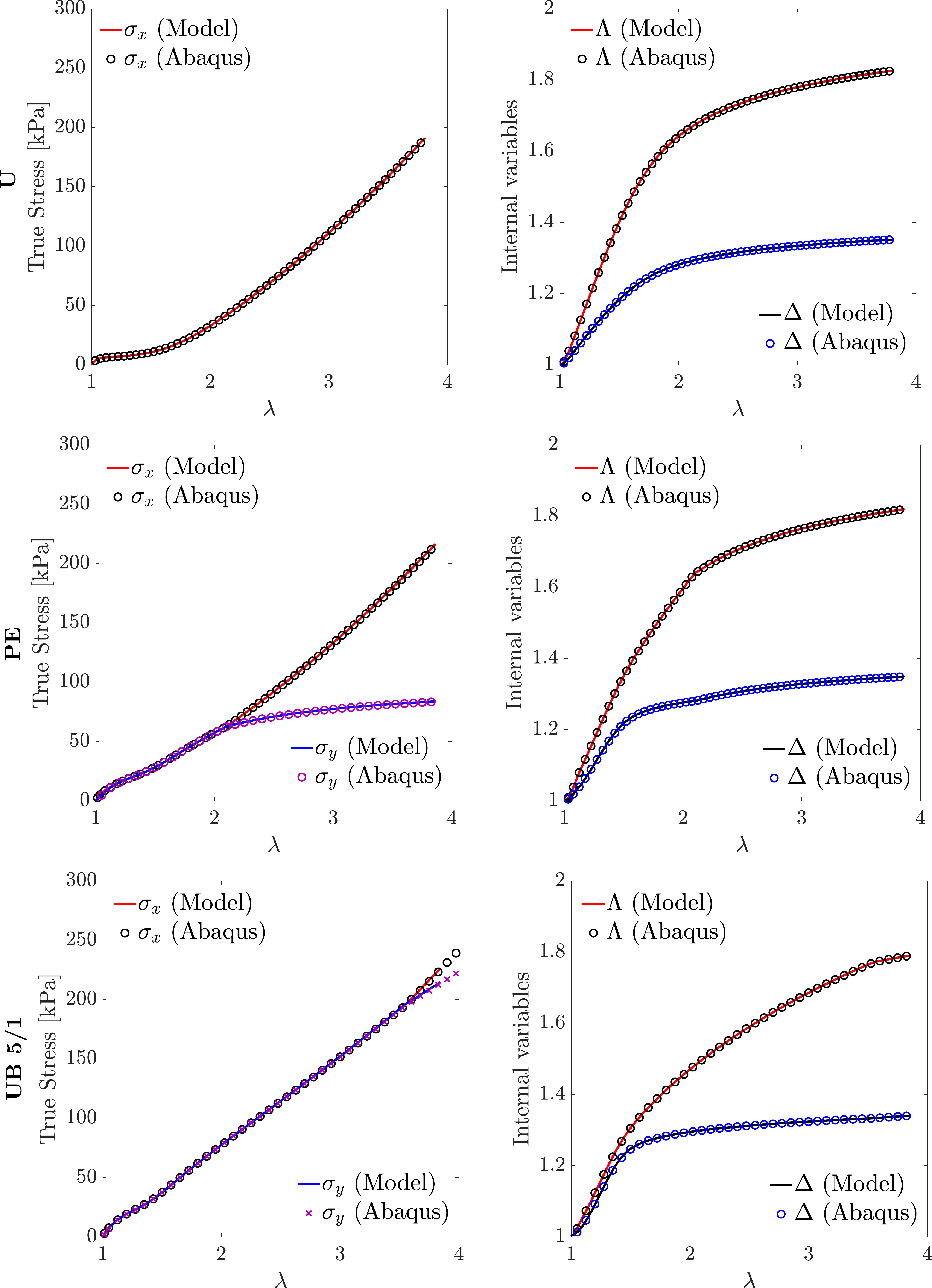}
\caption{Verification of the numerical implementation.  Each row represents a different loading mode described in Table \ref{tab:lm}.  The first column shows the true stress vs.\ stretch while the second column the state variables vs. stretch.  In each image, the numerical results are indicated by circles and they agree well with the model shown in the continuous line.  The figure continues below.\label{fig:verif}}
\end{figure}
\renewcommand{\thefigure}{\arabic{figure} (cont.)}
\setcounter{figure}{9}

\begin{figure}
\centering
\includegraphics[width=5 in]{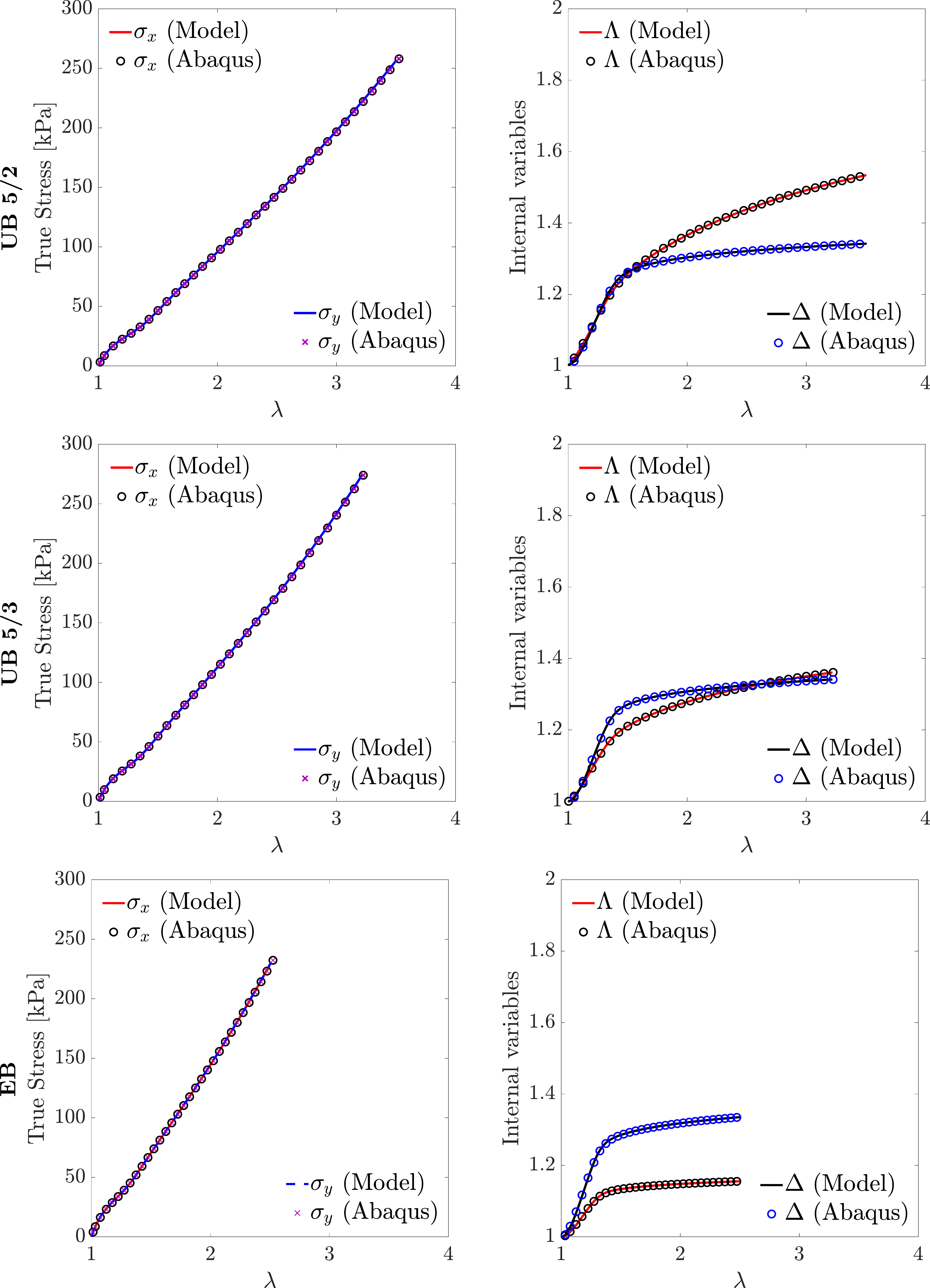}
\caption{Verification of the numerical implementation.  Each row represents a different loading mode described in Table \ref{tab:lm}.  The first column shows the true stress vs.\ stretch while the second column the state variables vs. stretch.  In each image, the numerical results are indicated by circles and they agree well with the model shown in the continuous line.}
\end{figure}
\renewcommand{\thefigure}{\arabic{figure}}

The reformulation and the implementation are verified against the uniaxial and biaxial extension tests described in Section \ref{sec:valid}.  We use the same values for the parameters, except we take a non-zero viscosity.  The implementation requires a bulk modulus and this is taken to be $10^4$ times the value of the shear modulus.  All the parameters are summarized in  Table \ref{tab:param}.

Figure \ref{fig:verif} compares the stress vs. stretch as well as the state variable vs. stretch for the various loading tests as obtained by the numerical implementation in \texttt{ABAQUS} with those obtained with the original model.  We see that the two sets of results agree very well, thereby verifying the implementation.

\subsection{Example}

\begin{figure}
\centering
\includegraphics[width=5in]{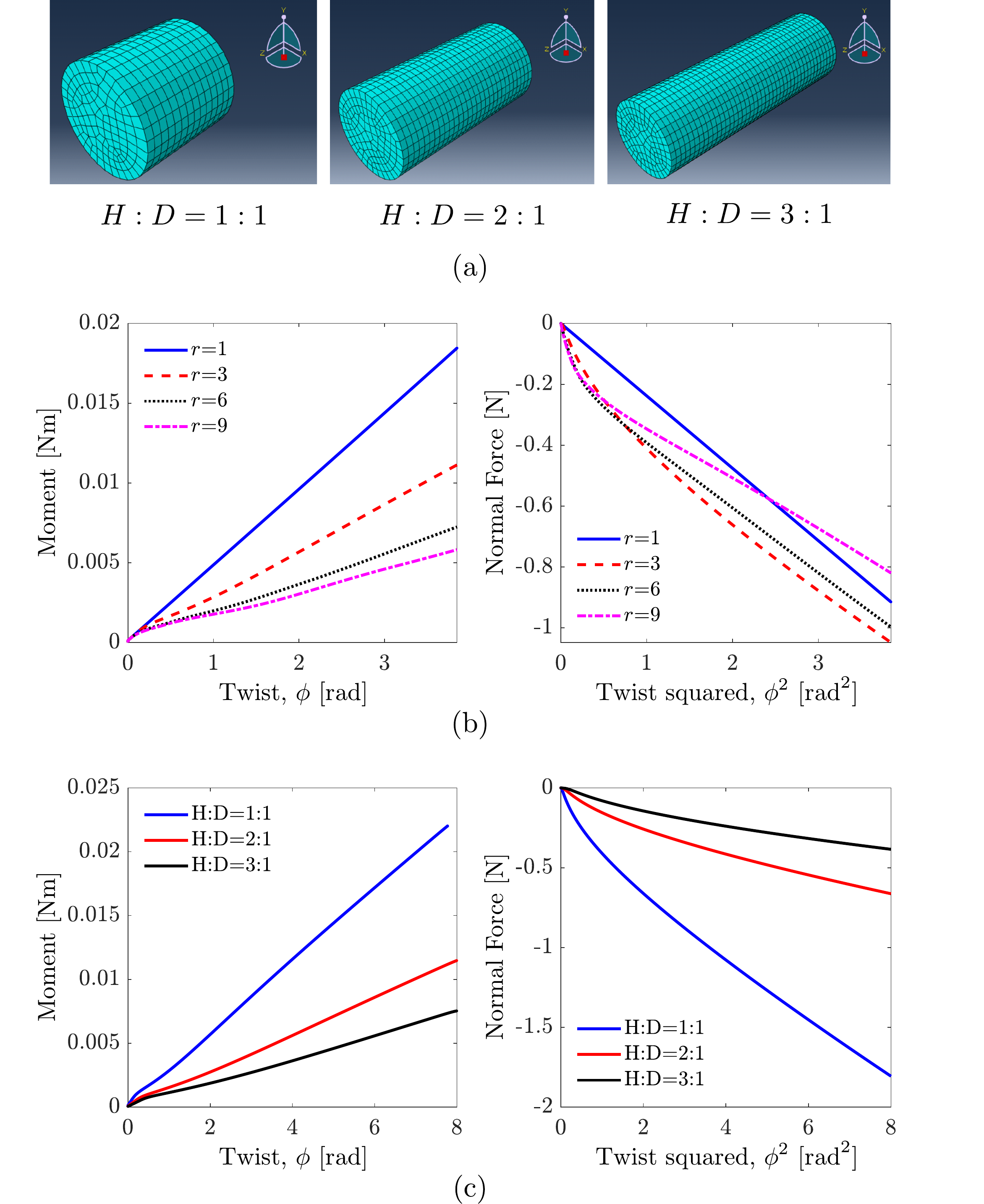}
\caption{Finite element calculations of torsion.  (a) The aspect ratios and the mesh.  (b) Moment and normal force as a function of twist for various values of the anisotropy parameter $r$.  $H/D=1$.  (c) Moment and normal force as a function of twist for various values of the aspect ratio $H/D$.  $r=3$. \label{fig:mf}}
\end{figure}

We now demonstrate the numerical implementation by studying the boundary value problem associated with torsion.  We consider cylinders of diameter $D$ and height $H$, and mesh them using C3D8H element in \texttt{ABAQUS}.  Specifically consider cylinders of diameter $D = 0.01$ and heights $H = 0.01, 0.02, 0.03$ ($H/D = 1, 2, 3$), and mesh them with $1652, 3509, 4859$ elements respectively as shown in Figure \ref{fig:mf}(a).  One face of the cylinder is held fixed while an angular velocity of $\omega = 0.0026$ rad/s is imposed on the other face.  The faces are constrained to be parallel to each other maintain a fixed distance (fixed height).

Figure \ref{fig:mf}(b),(c) shows the moment and normal force as a function of the twist for various values of anisotropy parameter $r$ (with $H/D$ fixed at 1) and $H/R$ (with $r$ fixed at 3).  We observe in Figure \ref{fig:mf}(b) that for the neo-Hookean material ($r=1$), the moment increases linearly with twist while the normal force increases quadratically with twist consistent with classical results.  The normal force is compressive, and its appearance reflects the Poynting effect.  When $r>1$, we see a nonlinear behavior for small twist in the moment vs. twist behavior reflecting the soft behavior of the LCE.  We see a corresponding non-quadratic behavior in the normal force vs. twist relation.  Interestingly, even though we have soft behavior, the normal force is higher for small twist for $r>1$ compared that at $r=1$.  At large twist, the normal force decreases with increasing $r$ and becomes quadratic.  Figure \ref{fig:mf}(c) studies the effect of aspect ratio H/D with $r$ fixed at 3.

\section{Conclusion} \label{sec:conc}

In this paper, we  present a macroscopic or engineering scale constitutive model of the behavior of I-PLCEs. The model implicitly accounts for the complex evolution of the domain patterns, and faithfully represents the observed response in multi-axial loading conditions. We validate the model against experiments and verify and demonstrate the numerical implementation.  The model and the implementation has been used to understand experimental observations on torsion-induced instabilities \cite{urayama_2023}, adhesion \cite{maghsoodi_2023} and Hertz contact \cite{terentjev_2023}.

We conclude by describing two interconnected areas of current work: non-isothermal processes and rate-dependence.  Most of this work is limited to isothermal processes and incorporates temperature through the material parameters.    Applications of LCE include actuation and shape morphing induced by change in temperature, and therefore non-isothermal processes are of interest.     Further, we only consider Newtonian viscosity and linear rate dependence.   Another area of potential application for LCEs is energy absorption at large deformations.  Therefore, high rate, large deformation behavior where linear viscosity and rate dependance may be inadequate is also of interest.   The framework can be extended to account for these phenomena, and this has been accomplished in  \ref{sec:dissipation}.   However, the implementation of this model will require us to extend the specification of the constitutive relation.  We can infer a temperature dependent free energy using experimental observations from uniaxial tensile response at various temperatures \cite{tajbakhsh_2001,azoug_2016}, as well as calorimetry \cite{warner_2003}.  It has also been shown that the viscosities change according to a time-temperature shift \cite{clarke_2002,azoug_2016}.  We can use these to extend the specification of the free energy functions and the viscosities.  However, the temperature and rate dependent study of large deformation is limited to monodomains \cite{wang_2022}, and remains an area of research in the polydomain materials.

\section*{Acknowledgement}  It is a pleasure to acknowledge the discussions with Kenji Urayama on the mechanics and physics of isotropic-genesis polydomain liquid crystal elastomers.  This work draws from the thesis of Victoria Lee at the California Institute of Technology.  We gratefully acknowledge the financial support of the Air Force Office of Scientific Research (FA9550-16-1-0566, KB and VL) and Office of Naval Research (N00014-18-1-2624, KB and AW).

\renewcommand\thesection{Appendix \Alph{section}}
\setcounter{section}{0}
\section{Consistent Jacobian} \label{sec:cj}

The consistent Jacobian ${\mathbb J}$ for the Jaumann rate of the Kirchoff stress defined in (\ref{eq: ddsdde}) is defined as such:
\begin{equation}
\begin{aligned}
    \mathbb J_{ijkl} &= -{2 \over 3} \mu J^{-5/3}\left[\left(F \tilde{G}^{-1}F^\top\right)_{ij}\delta_{kl} - {1 \over 3} \text{tr}\left(\tilde{G}^{-1} C\right) \delta_{ij}\delta_{kl} \right] \\
    &+ {\kappa \over J} \left[\delta_{ij} \delta_{kl} \right]\\
    &+ \mu J^{-5/3} \left[\left(F \tilde{G}^{-1} F^\top\right)_{lj}\delta_{ik} + \left(F \tilde{ G}^{-1} F ^\top\right)_{ik}\delta_{jl} \right]\\
    &+ \mu J^{-5/3} \left[-{2 \over 3} \left( F \tilde{G}^{-1} F^\top\right)_{kl} \delta_{ij} \right]\\
    &+ \mu J^{-5/3} \left[{1 \over |D|^2}\left(A_{ij} D_{kl}+D_{ij}A_{kl}\right) - {1 \over |D|^4} A_{mn} D_{mn} D_{ij} D_{kl}\right]
\end{aligned}
\end{equation}
where $A_{ij} = \left[\left(F \dot{\tilde{G}}^{-1} F^\top\right)_{ij} - {1 \over 3}\text{tr}\left(\dot{\tilde{G}}^{-1}C\right)\delta_{ij}\right]$ and $D$ the deformation rate.

\section{Arbitrary thermo-mechanical processes} \label{sec:dissipation}

We generalize the formulation in Section \ref{sec:model} to arbitrary thermo-mechanical processes.  We recall the balance of energy (first law of thermodynamics):
\begin{equation} \label{eq:1stlaw}
    \dot{U} - P \cdot \dot{F} -s +\nabla\cdot h = 0,
\end{equation}
where $U$ is the internal energy density (per unit reference volume), $P$  the first Piola-Kirchoff stress (nominal stress), $F$ the deformation gradient relative to the isotropic reference configuration, $s$  the body heat source density and $h$  the heat flux.  We also recall the Clausius-Duhem inequality (second law of thermodynamics):
\begin{equation} \label{eq:2ndlaw}
\dot{\eta} -\frac{s}{T} + \nabla \cdot \left(\frac{h}{T}\right) \geq 0,
\end{equation}
where $\eta$ and $T$ are the entropy density and temperature respectively.   We can use (\ref{eq:1stlaw}) to rewrite (\ref{eq:2ndlaw}) as the dissipation inequality:
\begin{equation} \label{eq: dissineq}
-\dot{W}-\dot{T}\eta + P \cdot \dot{F} - \frac{1}{T} h \cdot \nabla T  \geq 0
\end{equation}
where we introduce the Helmholtz free energy density $W=U-T \eta$.

We make the constitutive assumptions
\begin{equation}
\begin{aligned}
& W= W(F,\Lambda,\Delta, T), \quad h = h(\nabla T, T), \\
& \eta= \eta(F,\Lambda,\Delta, T), \quad
P = P^e (F,\Lambda,\Delta, T) + P^v(F,\dot{F},\Lambda,\Delta, T), \quad
\end{aligned}
\end{equation}
where we divide the stress as a sum of elastic and viscous stress.  Therefore, the dissipation inequality implies
\begin{equation} \label{eq: dissineq2}
\left(-\frac{\partial{W}}{\partial{F}} + P^e \right) \cdot \dot{F} + 
\left( - \frac{\partial{W}}{\partial{T}} - \eta \right)\dot{T} 
-\frac{\partial{W}}{\partial{\Lambda}} \dot{\Lambda} - 
\frac{\partial{W}}{\partial{\Delta}} \dot{\Delta} + P^v \cdot \dot{F} - \frac{1}{T} h \cdot \nabla T \geq 0.
\end{equation}
Assuming that arbitrary processes may be created by body force and body heat source density, we may argue as Coleman and Noll \cite{coleman_1963} that
\begin{equation}
P^e = \frac{\partial{W}}{\partial{F}}, \quad \eta = - \frac{\partial{W}}{\partial{T}}, \quad h \cdot \nabla T \le 0.
\end{equation}
Therefore, 
\begin{equation} \label{eq:dissineq3}
-\frac{\partial{W}}{\partial{\Lambda}} \dot{\Lambda} - \frac{\partial{W}}{\partial{\Delta}} \dot{\Delta} + P^v \cdot \dot{F}  \geq 0.
\end{equation}
It is convenient to use frame-indifference and to specify the viscous stress in terms with the Cauchy stress and rate of deformation $D$, i.e., $\sigma^v(D,F, \Lambda,\Delta, T)$ (instead of $P^v(F,\dot{F},\Lambda,\Delta, T)$) so that (\ref{eq:dissineq3}) can be rewritten
\begin{equation} \label{eq:dissineq4}
-\frac{\partial{W}}{\partial{\Lambda}} \dot{\Lambda} - \frac{\partial{W}}{\partial{\Delta}} \dot{\Delta} + \sigma^v \cdot D  \geq 0.
\end{equation}
Thus, $-\frac{\partial{W}}{\partial{\Lambda}}$ and $-\frac{\partial{W}}{\partial{\Delta}}$ are the force conjugates to the rate of change of the state variables.  therefore we identify them as the driving forces associated with the evolution of the state variables (\ref{eq:state1}) as in (\ref{eq:state1}) and postulate the kinetic relation
\begin{eqnarray}
\dot \Lambda = K_\Lambda (d_\Lambda, d_\Delta), \quad \quad
\dot \Delta = K_\Delta (d_\Lambda,d_\Delta)
\end{eqnarray}
subject to (\ref{eq:dissineq3}).  The specific choice (\ref{eq:state3}) along with Newtonian viscosity is a special case that we find is sufficient for our purposes.

We conclude by considering the general linear evolution laws following Leslie and Ericksen\cite{ericksen_1961,leslie_1968}  who derived the  hydrodynamic theory for nematic liquid crystals.  We postulate that the viscous stress and driving force for the evolution of state variables are linear in the rate.  In order to do so, recall that an LCE is incompressible, and therefore the rate of deformation tensor is purely deviatoric ($\text{tr } D = 0$).   Further, I-PLCE is isotropic.  
Therefore, we postulate 
\begin{align} 
d = N r \quad \text{where} \quad
d =  \begin{pmatrix} \sigma^v\\ d_\Lambda\\ d_\Delta \\ \end{pmatrix}, \quad
N = \begin{pmatrix}
\nu & \nu_{D\Lambda} & \nu_{D\Delta} \\
\nu_{\Lambda D} & \nu_\Lambda & \nu_{\Lambda\Delta}\\
\nu_{\Delta D} & \nu_{\Delta\Lambda} & \nu_\Delta\\
\end{pmatrix} 
r = \begin{pmatrix} D \\ \dot{\Lambda}\\ \dot{\Delta}\\ \end{pmatrix}
\end{align}
by accounting for material symmetry, and 
where $d$ is the driving force vector, $N$ is the viscosity matrix and $r$ is the rate vector.  
The dissipation inequality (\ref{eq:dissineq3}) may be written as
\begin{equation}
r^TN r \ge 0.
\end{equation}
It is common to assume that the left hand-side is associated with a dissipation potential which means that we can take viscosity matrix $N$ to be symmetric (Onsager reciprocity).  This implies that $N$ is symmetric, i.e.,  $\nu_{D\Lambda} = \nu_{\Lambda D}, \nu_{D\Delta} = \nu_{\Delta D},  \nu_{\Lambda\Delta} =  \nu_{\Delta\Lambda}$.  The dissipation inequality requires $N$ to be positive semidefinite.  This implies that we may have six linear viscosities that are independent up to the constraint that the viscosity matrix $N$ is positive.  The formulation in Section \ref{sec:model} (equations (\ref{eq:cauchy}) and (\ref{eq:state3})) is a special case where the off-diagonal terms are zero.

Finally, we can use the constitutive relations into the energy balance (\ref{eq:1stlaw}) to rewrite it as
\begin{equation} \label{eq:1stlaw2}
T \dot{\eta} = P^v \cdot \dot{F} + d_\Lambda \dot{\Lambda} + d_\Delta \dot{\Delta} + s - \nabla \cdot h.
\end{equation}

\newpage
\bibliographystyle{abbrv} %
\bibliography{lce_const_refs}

\end{document}